\documentclass[lettersize,journal]{IEEEtran}
\usepackage{amsmath,amsfonts}
\usepackage{xcolor}
\usepackage{algorithmic}
\usepackage{algorithm}
\usepackage{array}
\usepackage[caption=false]{subfig}
\usepackage{textcomp}
\usepackage{stfloats}
\usepackage{url}
\usepackage{verbatim}
\usepackage{nicefrac,xfrac}
\usepackage{graphicx}
\usepackage[noadjust]{cite}
\usepackage{mathtools}
\usepackage{bm,amsmath, amssymb}
\usepackage{amsmath}
\usepackage{amsfonts}
\usepackage{dsfont}
\usepackage{amssymb,amsmath}

\newtheorem{lemma}{Lemma}
\newtheorem{identity}{Identity}

\usepackage{flushend}

\begin{document}

\title{Low-Complexity Equalization of \\ Zak-OTFS in the Frequency Domain}

\author{Sandesh Rao Mattu$^*$, Nishant Mehrotra$^*$, Saif Khan Mohammed,\\ Venkatesh Khammammetti, and Robert Calderbank
\thanks{This work is supported by the National Science Foundation under grants 2342690 and 2148212, in part by funds from federal agency and industry partners as specified in the Resilient \& Intelligent NextG Systems (RINGS) program, by the Department of Science and Technology, Govt. of India under grant TPN-96226, and in part by the Air Force Office of Scientific Research under grants FA 8750-20-2-0504 and FA 9550-23-1-0249 \\ S. R. Mattu, N. Mehrotra, V. Khammammetti and R. Calderbank are with the Department of Electrical and Computer Engineering, Duke University, Durham, NC, 27708, USA (email: \{sandesh.mattu,~nishant.mehrotra,~venkatesh.khammammetti,~robert.calderbank\}@duke.edu). S. K. Mohammed is with Department of Electrical Engineering, Indian Institute of Technology Delhi, India. S. K. Mohammed is also associated with Bharti School of Telecom. Technology and Management (BSTTM), IIT Delhi (e-mail: saifkmohammed@gmail.com). \\$^*$ denotes equal contribution. \\ \noindent\textit{npj Wireless Technology, 2025}\\
\copyright 2025 Copyright held by the owner/author(s).\\
https://doi.org/XXXXXXX.XXXXXXX}\vspace{-5mm}}



\maketitle

\vspace{-3mm}

\begin{abstract}
4G/5G wireless standards use orthogonal frequency division multiplexing (OFDM) which is robust to frequency selectivity. Equalization is possible with a single tap filter, and low-complexity equalization makes OFDM an attractive physical layer. However the performance of OFDM degrades with mobility, since Doppler spreads introduce inter-carrier interference (ICI) between subcarriers and they are no longer orthogonal. Zak-transform based orthogonal time frequency space (Zak-OTFS) modulation has been shown to be robust to doubly selective channels. Zak-OTFS signals are formed in the delay-Doppler (DD) domain, converted to time domain (TD) for transmission and reception, then returned to the DD domain for processing. The received signal is a superposition of many attenuated copies since the doubly selective channel introduces delay and Doppler shifts. The received symbols are more difficult to equalize since they are subject to interference along both delay and Doppler axes. In this paper, we propose a new low-complexity method of equalizing Zak-OTFS in the frequency domain (FD). We derive the FD system model and show that it is unitarily equivalent to the DD system model. We show that the channel matrix in the FD is banded, making it possible to apply conjugate gradient methods to reduce the complexity of equalization. We show that complexity of FD equalization is linear in the dimension  of a Zak-OTFS frame. For comparison the complexity of naive MMSE equalization is cubic in the frame dimension. Through numerical simulations we show that FD equalization of Zak-OTFS achieves similar performance as equalization in DD domain.
\end{abstract}


\section{Introduction}
\label{sec:intro}

\IEEEPARstart{O}{rthogonal} frequency division multiplexing (OFDM) employs orthogonal subcarriers separated by a parameter called the subcarrier spacing. It is well-adapted to frequency selective channels since frequency selectivity converts the subcarriers to parallel flat-fading channels~\cite{tse2005fundamentals}.
In 4G/5G systems the carrier spacing is chosen so that inter-carrier interference (ICI) is of the same order as the underlying Gaussian noise. The subcarriers are effectively orthogonal, and the channel matrix is effectively diagonal. This makes the complexity of equalization very low; a one-tap equalizer divides each received symbol by the channel it experiences.
This 
defining
feature of OFDM has led to its widespread adoption.

As mobility increases (or carrier spacing decreases) the wireless channel becomes increasingly time-selective, and the subcarriers experience significant frequency shifts that vary from one subcarrier to another. 
ICI becomes more significant, the subcarriers are no longer effectively orthogonal, and the channel matrix is no longer effectively diagonal. It becomes increasingly difficult to accurately estimate the off-diagonal entries, and OFDM performance degrades with increasing Doppler spread (defined as twice the maximum absolute Doppler shift)~\cite{bitspaper1, bitspaper2}.


Doppler spread is directly proportional to the carrier frequency and the maximum absolute relative velocity between the transmitter and receiver. 
Spectrum in the sub-6 GHz range is already crowded, and increased demand has increased interest in higher carrier frequencies where a modest relative velocity can introduce large Doppler spreads.
Vertical applications such as communication with bullet trains also introduce large Doppler spreads. OFDM may only be able to satisfy these use cases by increasing carrier spacing (and reducing spectral efficiency).

Zak-transform based orthogonal time frequency space (Zak-OTFS) introduced in~\cite{bitspaper1, bitspaper2} is a modulation scheme that is robust to doubly selective channels. In Zak-OTFS, the information symbols are mounted in the delay-Doppler (DD) domain. The channel is also viewed in the DD domain. The received information symbols are simply the delay \& Doppler shifted and attenuated versions of the transmitted information symbols. 
This representation of the scattering environment changes at the speed of physics, which is much slower than the speed of wireless communication.

\begin{figure*}
    \centering
    \includegraphics[width=0.8\linewidth]{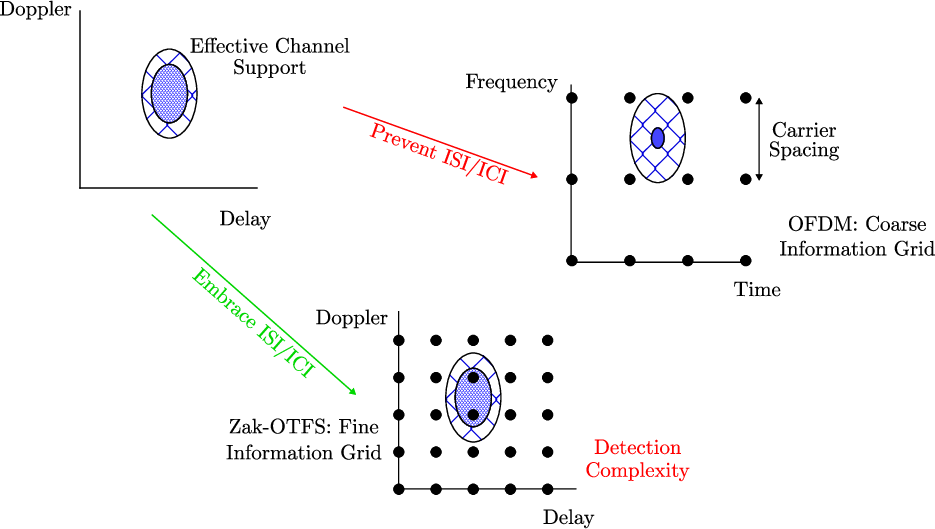}
    \caption{A pictorial representation of the difference between OFDM and Zak-OTFS. OFDM prevents ISI/ICI, that is the subcarrier spacing is adjusted to avoid interference, while Zak-OTFS embraces ISI/ICI. This results in poor resource utilization in OFDM but low equalization complexity while resource utilization is good in Zak-OTFS at the cost of equalization complexity.}
    \label{fig:prevent_vs_embrace}
\end{figure*}

Fig.~\ref{fig:prevent_vs_embrace} illustrates the fundamental difference between OFDM and Zak-OTFS. 
Consider a multi-user setup where many users are served in a slot with OFDM and Zak-OTFS. In OFDM, the subcarrier spacing is adjusted to cater to the worst Doppler spread across users, which means for all the other users, even if their Doppler spread is much smaller, the subcarrier spacing is still the same. This amounts to preventing or avoiding inter-symbol interference (ISI)/inter-carrier interference (ICI) which is illustrated in the OFDM grid points in Fig.~\ref{fig:prevent_vs_embrace} wherein the subcarrier spacing is chosen to be at least twice the Doppler spread. The benefit of avoiding ICI/ISI is the low equalization complexity. On the other hand, in Zak-OTFS, since the goal is not to choose grid points to avoid interference, we allow the information symbols to undergo ICI/ISI which results in higher equalization complexity. In a nutshell, while OFDM is designed to prevent interference, Zak-OTFS is designed to increase spectral efficiency by embracing interference. 

\begin{figure*}
    \subfloat[{OFDM}]{\includegraphics[width=0.49\linewidth]{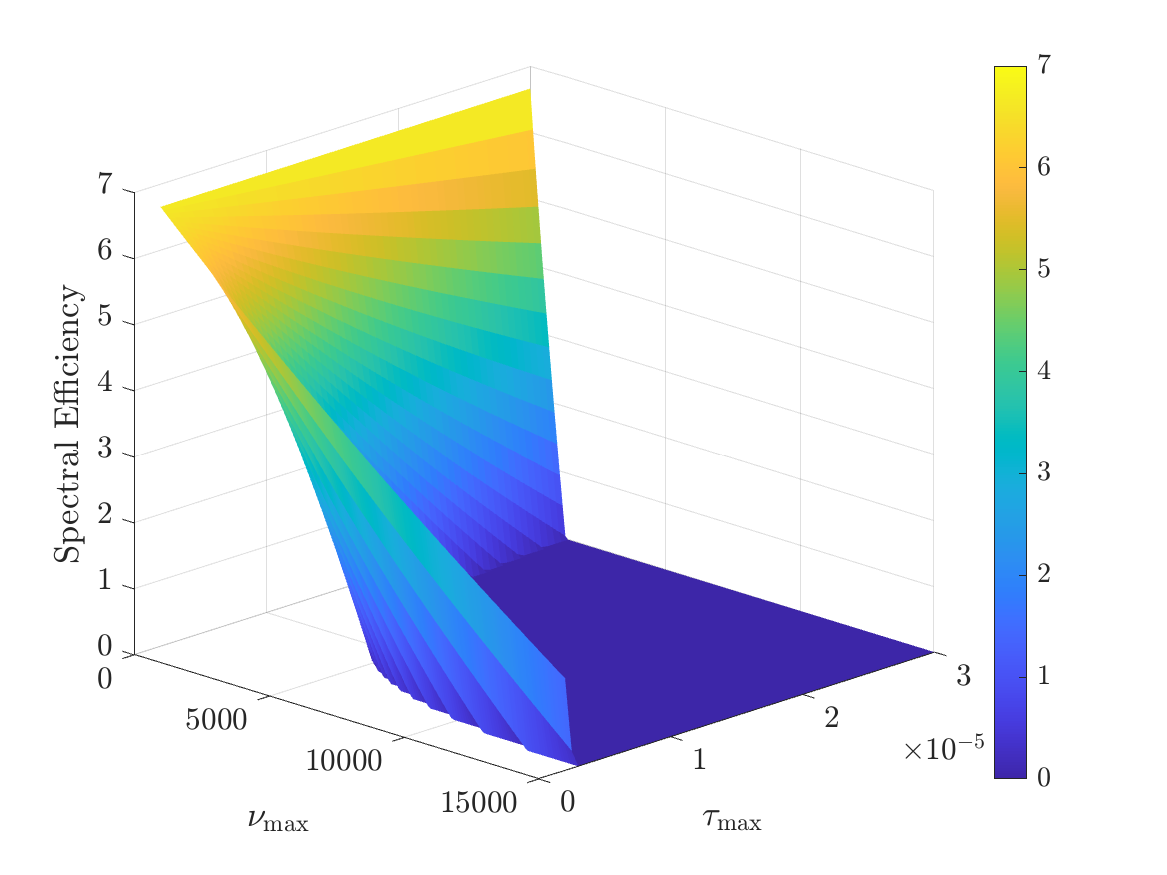}\label{fig:ofdm_se_surf}}
    \hfill
    \subfloat[{Zak-OTFS}]{\includegraphics[width=0.49\linewidth]{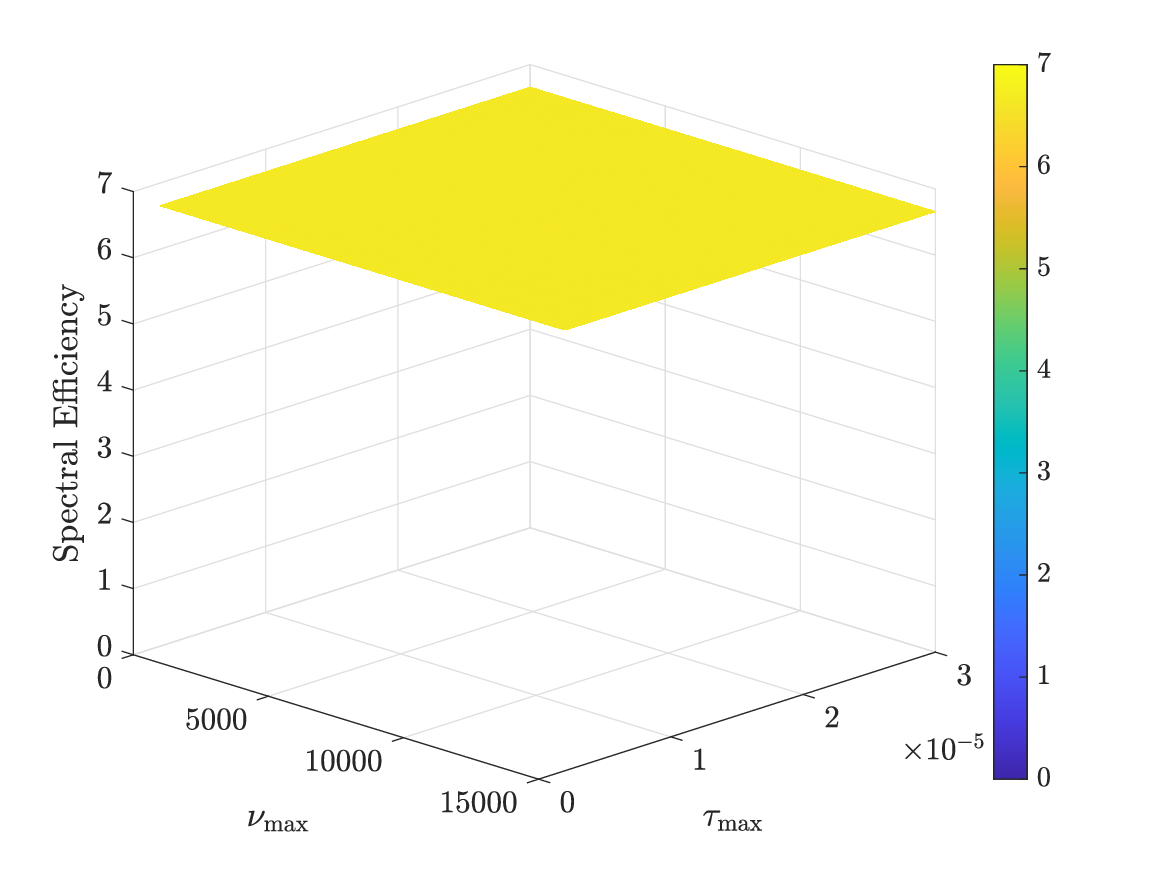}\label{fig:otfs_se_surf}}
    \hfill
    \caption{Spectral efficiency as a function of maximum delay and Doppler spread. OFDM spectral efficiency has a large variance and drops down to zero at high Doppler spread. Zak-OTFS has constant spectral efficiency as a function of maximum delay and Doppler spread.}
    \label{fig:se_surf}
\end{figure*}

\begin{table*}
    \centering
    \caption{Transceiver processing steps for FD equalization of Zak-OTFS assuming perfect channel knowledge.}
    {
    \setlength{\tabcolsep}{4.5pt}
    \renewcommand{\arraystretch}{1.5}
    \begin{tabular}{|p{6cm}|p{10cm}|}
         \hline
         \textbf{Transceiver Operation} & \textbf{Description} \\
         \hline
         Information symbol generation & Generate the $M \times N$ matrix of i.i.d. information symbols $\mathbf{X}[k_0,l_0]$, $0\leq k_0 \leq (M-1)$, $0\leq l_0 \leq (N-1)$, in the delay-Doppler (DD) domain \\
         \hline
         Convert to FD & Perform the Inverse Discrete Frequency Zak Transform (IDFZT) as per~\eqref{eq:zakofdm4} to transform the DD information symbols $\mathbf{X}[k_0,l_0]$ to the $MN \times 1$ FD vector $\mathbf{s}$ \\
         \hline
         Mask the FD vector & Zero out the first \& last $b$ entries in the FD vector $\mathbf{s}$ to generate $\mathbf{s}'$ according to~\eqref{eq:low_comp_eq2} \\
         \hline
         Channel propagation & Transmit the masked FD vector $\mathbf{s}'$ through a doubly-spread channel to receive FD vector $\mathbf{r}'$ as per~\eqref{eq:low_comp_eq3}  \\
         \hline
         Low-complexity equalization & Perform low-complexity equalization on $\mathbf{r}'$ to obtain FD vector $\tilde{\mathbf{s}}$ via conjugate gradient method in Algorithm~\ref{alg:conj_grad} \\
         \hline
         Information symbol detection & Perform the DFZT on $\tilde{\mathbf{s}}$ and detect information symbols in the DD domain via~\eqref{eq:low_comp_eq4} \\
         \hline
    \end{tabular}
    }
    \label{tab:tx_rx_proc}
\end{table*}

This is illustrated in Fig.~\ref{fig:se_surf} where we plot the spectral efficiency (in bits/s/Hz) of OFDM and the proposed Zak-OTFS as a function of maximum delay spread $(\tau_{\max})$ and Doppler spread $(\nu_{\max})$ for a frame bandwidth of $720$ kHz and frame duration of $1$ ms~\cite{khan2025waveform}. It is seen that the spectral efficiency of OFDM varies between 7 and 0. The spectral efficiency being zero implies that the corresponding $(\nu_{\max}, \tau_{\max})$ pair is not supported by OFDM. On the other hand, the spectral efficiency of Zak-OTFS remains constant within the considered $\nu_{\max}$ and $\tau_{\max}$ range.

For a channel with $P$ resolvable paths in the DD domain, $P$ (delay \& Doppler shifted and attenuated) copies of each transmitted symbol are received. This introduces interference along both delay and Doppler axes. The effective channel matrix is no longer sparse and does not have a structure that could be exploited to equalize the received information symbols at low complexity. The current implementations of Zak-OTFS \cite{Aug2024paper, preamblepaper} all use the linear minimum mean square error (LMMSE) matrix \cite{tse2005fundamentals} to equalize the received information symbols. However, for large frame sizes the LMMSE matrix computation becomes impractical since its complexity is cubic in the frame size owing to 
the need for matrix inversion.

The key contributions of the paper are listed as follows.
\begin{itemize}
    \item We propose a low-complexity equalization scheme for Zak-OTFS, where equalization is carried out in the \emph{frequency-domain} (FD)\footnote{A parallel development can be done in the time-domain (TD) as well. Although equalization of channel in the TD may be beneficial in time-selective channels, channel matrix is not sparse with frequency-selective channels. Since this paper deals with fractional DD channel, the TD equalization incurs high complexity owing to the TD channel matrix being dense.\label{footnote:td_matrix}} instead of the DD domain. Specifically, the information symbols in the DD domain are converted to the FD using the inverse discrete frequency Zak transform (IDFZT)~\cite{otfs_book} at the transmitter. These symbols are transmitted through a doubly selective channel with \textit{fractional} DD values. 
    \item We derive a system model in the FD that relates the transmitted FD information symbols to the received FD information symbols. We show that this matrix is modulo banded diagonal (see Fig.~\ref{fig:mod_banded}).
    \item Using the null space of the IDFZT matrix, we reduce the channel matrix to a banded diagonal that admits a low-complexity conjugate gradient method (CGM)~\cite{liu2020energy} for equalization.
    \item We show that the complexity of the proposed low-complexity equalization scheme is linear in the frame size compared to the cubic complexity of LMMSE.
    \item Finally, through extensive numerical simulations, we show that FD equalization of Zak-OTFS achieves similar performance as equalization in DD in both perfect and estimated channel knowledge scenarios across various choices of pulse shaping filters~\cite{das2025gaussian}.
\end{itemize}

A preliminary version of this method was proposed in~\cite{zak_ofdm_letter}. The main differences between~\cite{zak_ofdm_letter} and this paper are as follows. In this paper we derive the matrix form of the input-output (I/O) relation in the FD, which is not present in~\cite{zak_ofdm_letter}. Next, to reduce the modulo-banded channel matrix in the FD to a banded structure, we mount symbols on the null space of the transform matrix, which is not considered in~\cite{zak_ofdm_letter}. We also consider conjugate gradient method for equalization in this paper which has complexity linear in frame size; in contrast the complexity of the method in~\cite{zak_ofdm_letter} is quadratic. Lastly, we study the effect of estimated channel state information (CSI) with various pulse shapes, which is not considered in~\cite{zak_ofdm_letter}.

A high level description of the transceiver processing steps of the proposed FD equalization of Zak-OTFS is presented in Table~\ref{tab:tx_rx_proc}. Detailed description of each step is provided in the following Sections.

\textit{Notation:} $x$ denotes a complex scalar, $\mathbf{x}$ denotes a vector with $n$th entry $\mathbf{x}[n]$, and $\mathbf{X}$ denotes a matrix with $(n,m)$th entry $\mathbf{X}[n,m]$. $(\cdot)^{\ast}$ denotes complex conjugate, $(\cdot)^{\mathsf{T}}$ denotes transpose, $(\cdot)^{\mathsf{H}}$ denotes complex conjugate transpose, and $(\cdot)^{-1}$ denotes the matrix inverse. $\mathbb{Z}$ denotes the set of integers and $\mathbb{C}$ denotes the set of complex numbers. $\delta[\cdot]$ denotes the Kronecker delta function. $\Vert\cdot\Vert_2^2$ denotes the 2-norm of a vector or Frobenius norm of a matrix. $a \equiv 0 \bmod N \implies a = cN$ for some $c\in \mathbb{Z}$. $\otimes$ denotes the Kronecker product. $\mathds{1}$ denotes the indicator function. $\mathsf{null}(\mathbf{A})$ denotes the null space of the matrix $\mathbf{A}$, $\mathsf{vec}(\cdot)$ denotes column-wise vectorization, and $\mathsf{diag}(\mathbf{a})$ denotes a diagonal matrix with $\mathbf{a}$ as the diagonal. $\mathbf{A}\succ 0$ implies that the matrix $\mathbf{A}$ is positive definite. $\nabla$ denotes the vector gradient. $\mathcal{O}(\cdot)$ denotes the complexity order. $\lceil \cdot \rceil$ denotes the ceil operation.

\section{Prior work}
\label{sec:prior_work}

\begin{table*}
    \centering
    \caption{Characteristics of different methods. $b$ indicates the spread width and $k$ indicates the number of iterations of conjugate gradient method.} 
    {
    \setlength{\tabcolsep}{4.5pt}
    \renewcommand{\arraystretch}{1.25}
    \begin{tabular}{|c|c|c|c|c|}
         \hline
         Method & Mobility & Channel Type & Predictable? & Complexity \\
         \hline
         FD equalization of Zak-OTFS (Ours) & Low \& High & Fractional DD & \checkmark & $\mathcal{O}(bkMN)$ \\
         Zak-OTFS (Ours) & Low \& High & Fractional DD & \checkmark & $\mathcal{O}(kMN^2)$ \\
         FD equalization of Zak-OTFS~\cite{zak_ofdm_letter} & Low \& High & Fractional DD & \checkmark & $\mathcal{O}(M^2N^2)$ \\
         Zak-OTFS-OGD~\cite{Zhang2025_lowcompl_zakotfs} & Low & Integer DD & $\times$ & $\mathcal{O}(M^2N^2)$ \\
         Zak-OTFS~\cite{bitspaper1,bitspaper2,otfs_book,gopalam2024zak} & Low \& High & Fractional DD & \checkmark & $\mathcal{O}(M^3N^3)$ \\
         MC-OTFS~\cite{Hadani2017,otfs_book,Chockalingam2020,gopalam2024new} & Low \& High & Fractional DD & $\times$ & $\mathcal{O}(MN\log MN)$ \\
         OFDM & Low & Fractional DD & $\times$ & $\mathcal{O}(MN)$ \\
         OCDM~\cite{ocdm_1, ocdm_2} &  Low \& High & Integer delay, Fractional Doppler & $\times$ & $\mathcal{O}(MN \log MN)$ \\
         AFDM~\cite{afdm_1, afdm_2, otfs_afdm_1, otfs_afdm_2} & Low \& High & Integer delay, Fractional Doppler & $\times$ & $\mathcal{O}((MN-2b)k)$ \\
         ODDM~\cite{oddm} & Low \& High & Integer delay, Fractional Doppler & $\times$ & $\mathcal{O}(kMN)$ \\
         \hline
    \end{tabular}
    }
    \label{tab:prior_work}
\end{table*}

There are several modulation schemes proposed in the literature~\cite{ocdm_1, Zhang2025_lowcompl_zakotfs, ocdm_2, afdm_1, afdm_2, oddm, otfs_afdm_1, otfs_afdm_2, Hadani2017, otfs_book, Chockalingam2020}. For obvious reasons, we do not compare bit-error performance against all the schemes proposed in literature. However, we quantitatively differentiate Zak-OTFS from these schemes in the following.

The modulation schemes in literature are presented in Table~\ref{tab:prior_work}. Multicarrier-OTFS (MC-OTFS) approaches in~\cite{Hadani2017,otfs_book,Chockalingam2020} are not ``predictable" and the channel response from a pilot at a given location cannot be used to predict the channel response at another location~\cite{bitspaper2}. OFDM does not work under high mobility scenarios as described earlier and is not predictable. Zak-OTFS-OGD proposed in~\cite{Zhang2025_lowcompl_zakotfs} does not work with fractional DD channel and therefore is not practical. Orthogonal chirp domain modulation (OCDM)~\cite{ocdm_1, ocdm_2} are proposed with integer delay and fractional Doppler values and is not a true representation of the channel. A generalized version of OCDM is affine frequency domain multiplexing (AFDM)~\cite{afdm_1, afdm_2, otfs_afdm_1, otfs_afdm_2}. Again, the system model is proposed only for integer delay shifts and fractional Doppler shifts. Orthogonal delay-Doppler multiplexing (ODDM)~\cite{oddm} is yet another modulation scheme that works in the DD domain. However, like the previous schemes, the system model is proposed only for integer delay values. Further OCDM, AFDM, and ODDM all require cyclic prefix (CP) which Zak-OTFS does not. For channel estimation, all these methods make use of the embedded pilot as opposed to the spread pilot~\cite{Aug2024paper, preamblepaper} or even data symbols~\cite{mattu2025differential} in Zak-OTFS which incurs no spectral efficiency loss. Equalization of Zak-OTFS in the DD domain~\cite{bitspaper1,bitspaper2,otfs_book} incurs cubic complexity in the frame size owing to matrix inversion in the LMMSE matrix computation. FD equalization of Zak-OTFS presented in the preliminary version~\cite{zak_ofdm_letter} incurs quadratic complexity in the frame size. The approach presented in this paper has complexity linear in the frame size. This follows from reduction of the channel matrix to banded structure and exploiting this structure in the CGM. In the DD domain where the Zak-OTFS channel matrix does not have a defined structure, equalization via CGM allows for reduced complexity.

Among the modulation schemes, we pick AFDM for performance comparison (see Sec.~\ref{subsec:afdm_vs_zak}), since the system model can be extended for fractional DD values and also since it is the generalized version of OCDM. Further, AFDM and Zak-OTFS are the only two schemes which are added into the 6G standards as study items~\cite{6g_study_item}.

\textit{Remark:} There are many methods in the literature for MC-OTFS (for example~\cite{zhang2021adaptive, chong2025cross, yuan2021data, mishra2021otfs}) that aim to perform low-complexity equalization in different domains (including the time and/or frequency domain). These methods do not directly apply to Zak-OTFS. This is because MC-OTFS and Zak-OTFS are not related through unitary transforms (for a detailed discussion, see~\cite[Figs. 14 \& 15]{bitspaper2}). If it were, it should have been possible to move from one system model to another through unitary matrices. This also explains why some of the salient features in Zak-OTFS like predictability is missing in MC-OTFS. Hence, all the low-complexity methods for equalization proposed for MC-OTFS \textit{cannot} be applied in the context of Zak-OTFS.


Before we describe the FD perspective of Zak-OTFS, we present a few prelimaries that would be used throughout the paper.

\section{Preliminaries}
\label{sec:prelim}

\subsection{Useful Identities}
\label{subsec:prelim_identities}

\begin{identity}[\cite{murty2017evaluation}]
    \label{idty:sumrootsofunity}
    The sum of all $N$th roots of unity satisfies:
    \begin{align*}
        \sum_{n=0}^{N-1}e^{\frac{j2\pi}{N}kn} = \begin{cases}
        N \quad \text{if } \ k \equiv 0 \bmod{N} \\
        0 \quad \ \text{otherwise}
        \end{cases}.
    \end{align*}
\end{identity}

\subsection{Zak-OTFS System Model in Discrete Time}
\label{subsec:prelim_zak_otfs_time}

\subsubsection{Transmitter}

Consider an $M\times N$ delay-Doppler (DD) grid with $M$ delay bins (indexed by $k$) and $N$ Doppler bins (indexed by $l$), corresponding to width $\tau_p$ along delay and width $\nu_p$ along Doppler with $\tau_p \nu_p = 1$. Given $0\leq k_0 \leq (M-1),~0\leq l_0 \leq (N-1)$, the basis element corresponding to the $(k_0,l_0)$th DD bin is $\sum_{n,m\in\mathbb{Z}} e^{j\frac{2\pi}{N} ln} \delta[k-k_0-nM]\delta[l-l_0-mN]$. The corresponding $MN$-length discrete time realization is given by the inverse discrete Zak transform~\cite{dzt,otfs_book}:
\begin{equation}
    \label{eq:zakotfs1a}
    \mathbf{p}_{(k_0,l_0)}[n] = \frac{1}{\sqrt{N}} \sum_{d \in \mathbb{Z}} e^{j\frac{2\pi}{N} d l_0} \delta[n-k_0-dM],
\end{equation}
where $0\leq n \leq (MN-1)$, which 
we refer to as a time domain (TD) \textit{pulsone}. We refer the interested reader to~\cite{bitspaper1,bitspaper2} for detailed discussions regarding the pulsone basis.


The pulsone basis is orthonormal, and $MN$ DD domain information symbols $\mathbf{X}[k_0,l_0]  \in \mathbb{C}^{M\times N}$, $0\leq k_0 \leq (M-1),~0\leq l_0 \leq (N-1)$, can be transmitted by modulating different pulsones as:
\begin{equation}
    \label{eq:zakotfs2}
    \mathbf{x}[n] = \sum_{k_0=0}^{M-1} \sum_{l_0=0}^{N-1} \mathbf{X}[k_0,l_0] \mathbf{p}_{(k_0,l_0)}[n] \in \mathbb{C}^{MN\times 1}.
\end{equation}

Note that $\mathbf{p}_{(k_0,l_0)}[n]$, and hence $\mathbf{x}[n]$, is periodic with period $MN$. Therefore, $\mathbf{x}[n]$ can also be represented in terms of a time domain orthonormal basis $\boldsymbol{\phi}_{i}[n]$ $\in \mathbb{C}^{MN \times 1}$ that converts symbols in the transfer domain to time domain as:
\begin{align}
    \label{eq:zakotfs3}
    \mathbf{x}[n] &= \sum_{i = 0}^{MN-1} \mathbf{s}[i] \boldsymbol{\phi}_{i}[n],
\end{align}
\begin{figure*}
    \centering
    \includegraphics[width=\linewidth]{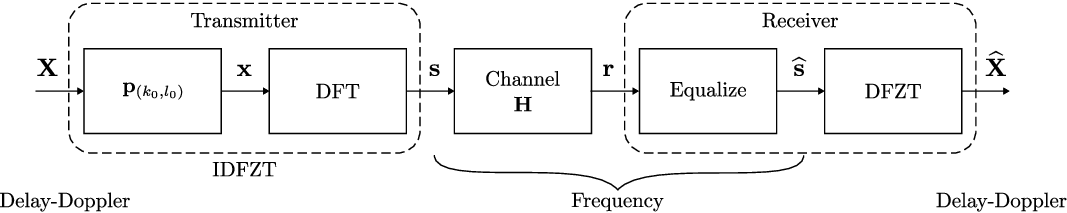}
        \caption{Block diagram of FD input-output relation of Zak-OTFS.}
    \label{fig:zak_ofdm_block_diag}
\end{figure*}
where $\mathbf{s}[i] \in \mathbb{C}^{MN \times 1}$, for $0 \leq i \leq (MN-1)$, is a vector in the transfer domain given by:
\begin{align}
    \label{eq:zakotfs4}
    \mathbf{s}[i] &= \sum_{n=0}^{MN-1} \boldsymbol{\phi}_{i}^{*}[n] \mathbf{x}[n].
\end{align}

\subsubsection{Receiver}

The (noiseless) received time domain signal is given by:
\begin{align}
    \label{eq:sys_model1}
    \mathbf{y}[n] &= \sum_{k, l \in \mathbb{Z}} \mathbf{h}_{\mathrm{eff}}[k, l] \mathbf{x}[n-k]e^{\frac{j2\pi}{MN}l(n-k)}  \in \mathbb{C}^{MN\times 1},
\end{align}
where $\mathbf{h}_{\mathrm{eff}}[k, l] \in \mathbb{C}^{\mathbb{Z}\times \mathbb{Z}}$ represents the effective channel in the DD domain. 

Substituting $k = \bar{k} + pMN$, $0 \leq \bar{k} \leq (MN-1)$, $p \in \mathbb{Z}$, $l = \bar{l} + qMN$, $0 \leq \bar{l} \leq (MN-1)$, $q \in \mathbb{Z}$:
\begin{align}
    \label{eq:sys_model2}
    \mathbf{y}[n] &= \sum_{\bar{k}=0}^{MN-1} \sum_{\bar{l}=0}^{MN-1} \sum_{p, q \in \mathbb{Z}} \mathbf{h}_{\mathrm{eff}}[\bar{k}+pMN, \bar{l}+qMN] \nonumber \\ &~~~~~~~\mathbf{x}[n-\bar{k}-pMN]e^{\frac{j2\pi}{MN}(\bar{l}+qMN)(n-\bar{k}-pMN)}.
\end{align}

Since $\mathbf{x}[n]$ is $MN$-periodic,~\eqref{eq:sys_model2} can be simplified as:
\begin{align}
    \label{eq:sys_model3}
    \mathbf{y}[n] &= \sum_{\bar{k}=0}^{MN-1} \sum_{\bar{l}=0}^{MN-1} \underbrace{\sum_{p, q \in \mathbb{Z}} \mathbf{h}_{\mathrm{eff}}[\bar{k}+pMN, \bar{l}+qMN]}_{\mathbf{h}[\bar{k},\bar{l}]} \times \nonumber \\ &~~~~~~~~~~~~~~~~~~~\mathbf{x}[n-\bar{k}]e^{\frac{j2\pi}{MN}\bar{l}(n-\bar{k})}.
\end{align}
where $\mathbf{h}[\bar{k},\bar{l}] \in \mathbb{C}^{MN\times MN}$ denotes the $MN$-periodized version of the effective channel in the DD domain. Therefore,
\begin{align}
    \label{eq:sys_model4a}
    \mathbf{y}[n] &= \sum_{\bar{k}=0}^{MN-1} \sum_{\bar{l}=0}^{MN-1} \mathbf{h}[\bar{k},\bar{l}] \mathbf{x}[n-\bar{k}]e^{\frac{j2\pi}{MN}\bar{l}(n-\bar{k})}.
\end{align}

Hence,
\begin{align}
    \label{eq:sys_model4}
    \mathbf{y}[n] &= \sum_{\bar{k}=0}^{MN-1} \sum_{\bar{l}=0}^{MN-1} \mathbf{h}[\bar{k},\bar{l}] e^{\frac{j2\pi}{MN}\bar{l}(n-\bar{k})} \times \nonumber \\ &~~~~\sum_{i = 0}^{MN-1} \mathbf{s}[i] \boldsymbol{\phi}_{i}[n-\bar{k}].
\end{align}

Projecting $\mathbf{y}[n]$ on the basis $\boldsymbol{\phi}_{f}[n] \in \mathbb{C}^{MN\times 1}$, $0 \leq f \leq (MN-1)$:
\begin{align}
    \mathbf{r}[f] &= \sum_{n=0}^{MN-1} \boldsymbol{\phi}_{f}^{*}[n] \mathbf{y}[n] \nonumber \\
    &= \sum_{i = 0}^{MN-1} \mathbf{s}[i] \bigg(\sum_{\bar{k}=0}^{MN-1} \sum_{\bar{l}=0}^{MN-1} e^{-\frac{j2\pi}{MN}\bar{l}\bar{k}} \mathbf{h}[\bar{k},\bar{l}] \times \nonumber \\ &~~~\sum_{n=0}^{MN-1} \boldsymbol{\phi}_{f}^{*}[n] \boldsymbol{\phi}_{i}[n-\bar{k}] e^{\frac{j2\pi}{MN}\bar{l}n}\bigg) \nonumber 
\end{align}
\begin{align}
\label{eq:sys_model5}
    = \sum_{i = 0}^{MN-1} \mathbf{s}[i] \mathbf{H}[f,i],
\end{align}
where $\mathbf{r}[f] \in \mathbb{C}^{MN\times 1}$ is in the transfer domain.

\subsection{Vectorized System Model}
\label{subsec:prelim_zak_otfs_vec}

Vectorizing~\eqref{eq:sys_model5} and including additive noise at the receiver, we obtain the system model: 
\begin{align}
    \label{eq:sys_model6}
    \mathbf{r} = \mathbf{H} \mathbf{s} + \mathbf{w},
\end{align}
where $\mathbf{r}, \mathbf{s}, \mathbf{w} \in \mathbb{C}^{MN\times 1}$ are all in the transfer domain and $\mathbf{H} \in \mathbb{C}^{MN \times MN}$ is the channel matrix in the transfer domain.

In the next Section, we describe the FD perspective of Zak-OTFS by appropriately choosing $\boldsymbol{\phi}[n]$.

\section{A FD Perspective of Zak-OTFS}
\label{sec:zak_ofdm}

In this Section, we study the FD input-output (I/O) relation of Zak-OTFS. Fig.~\ref{fig:zak_ofdm_block_diag} shows the block diagram of the FD input-output relation of Zak-OTFS.

\subsection{Transmitter Processing}
\label{subsec:zak_ofdm_tx}

To derive the FD input-output relation of Zak-OTFS, we use the inverse discrete Fourier transform (IDFT) basis:
\begin{align}
    \label{eq:zakofdm1}
    \boldsymbol{\phi}_{i}[n] &= \frac{1}{\sqrt{MN}} e^{j\frac{2\pi}{MN} i n}.
\end{align}
where the FD vector $\mathbf{s}[i] \in \mathbb{C}^{MN\times 1}$, for $0 \leq i \leq (MN-1)$, is given by:
\begin{align}
    \label{eq:zakofdm2}
    \mathbf{s}[i] &= \sum_{n=0}^{MN-1} \boldsymbol{\phi}_{i}^{*}[n] \mathbf{x}[n] \nonumber \\
    &= \frac{1}{\sqrt{MN}} \sum_{n=0}^{MN-1} e^{-j\frac{2\pi}{MN} i n} \mathbf{x}[n] \nonumber
    \end{align}
    \begin{align}
    &= \frac{1}{\sqrt{MN}} \sum_{k_0=0}^{M-1} \sum_{l_0=0}^{N-1} \mathbf{X}[k_0,l_0] \sum_{n=0}^{MN-1} e^{-j\frac{2\pi}{MN} i n} \mathbf{p}_{(k_0,l_0)}[n] \nonumber \\
    &= \frac{1}{\sqrt{M}N} \sum_{k_0=0}^{M-1} \sum_{l_0=0}^{N-1} \mathbf{X}[k_0,l_0] \sum_{d \in \mathbb{Z}} e^{j\frac{2\pi}{N} d l_0} \times \nonumber \\ &~~~~~~~~~~~~\sum_{n=0}^{MN-1} e^{-j\frac{2\pi}{MN} i n} \delta[n-k_0-dM].
\end{align}

Note that since $0\leq n \leq (MN-1)$, the delta function $\delta[n-k_0-dM]$ implies that $0 \leq d \leq (N-1)$ for which $n = k_0+dM$. Therefore,
\begin{align}
    \label{eq:zakofdm3}
    \mathbf{s}[i] &= \frac{1}{\sqrt{M}N} \sum_{k_0=0}^{M-1} \sum_{l_0=0}^{N-1} \mathbf{X}[k_0,l_0] \sum_{d = 0}^{N-1} e^{j\frac{2\pi}{N} d l_0} e^{-j\frac{2\pi}{MN} i (k_0+dM)} \nonumber \\
    &= \frac{1}{\sqrt{M}} \sum_{k_0=0}^{M-1} \sum_{l_0=0}^{N-1} \mathbf{X}[k_0,l_0] e^{-j\frac{2\pi}{MN} i k_0} \frac{1}{N} \sum_{d = 0}^{N-1} e^{j\frac{2\pi}{N} d (l_0-i)}.
\end{align}

From Identity~\ref{idty:sumrootsofunity}, the inner summation over $d$ vanishes unless $l_0 \equiv i \bmod N$, when it takes the value $N$. Therefore,
\begin{align}
    \label{eq:zakofdm4}
    \mathbf{s}[i] &= \frac{1}{\sqrt{M}} \sum_{k_0=0}^{M-1} \sum_{l_0=0}^{N-1} \mathbf{X}[k_0,l_0] e^{-j\frac{2\pi}{MN} i k_0} \mathds{1}\{l_0 \equiv i \bmod N\} \nonumber \\
    &= \frac{1}{\sqrt{M}} \sum_{k_0=0}^{M-1} \mathbf{X}[k_0,i \bmod N] e^{-j\frac{2\pi}{MN} i k_0}.
\end{align}

The above relation is called the \emph{inverse discrete frequency Zak transform} (IDFZT)~\cite[Chapter~8]{otfs_book}.

\subsection{Receiver Processing}
\label{subsec:zak_ofdm_rx}

The inner summation over $n$ in~\eqref{eq:sys_model5} for the discrete Fourier transform (DFT) basis in~\eqref{eq:zakofdm1} simplifies to:
\begin{align}
    \label{eq:zak_ofdm5}
    \sum_{n=0}^{MN-1} &\boldsymbol{\phi}_{f}^{*}[n] \boldsymbol{\phi}_{i}[n-\bar{k}] e^{\frac{j2\pi}{MN}\bar{l}n} \nonumber \\ &= \frac{1}{MN} \sum_{n=0}^{MN-1} e^{-j\frac{2\pi}{MN} f n} e^{j\frac{2\pi}{MN} i (n-\bar{k})} e^{\frac{j2\pi}{MN}\bar{l}n} \nonumber \\
    &= \frac{1}{MN} e^{-j\frac{2\pi}{MN} i \bar{k}} \sum_{n=0}^{MN-1} e^{j\frac{2\pi}{MN} (i-f+\bar{l}) n}.
\end{align} 

From Identity~\ref{idty:sumrootsofunity}, the inner summation over $n$ vanishes unless $\bar{l} \equiv (f-i) \bmod MN$, when it takes the value $MN$. Therefore,
\begin{align}
    \label{eq:zak_ofdm6}
    \mathbf{H}[f,i] &= \sum_{\bar{k}=0}^{MN-1} \mathbf{h}[\bar{k},(f-i) \bmod MN] e^{-j\frac{2\pi}{MN}f\bar{k}}.
\end{align}

\subsection{Unitary Equivalence of FD and DD I/O Relations}
\label{subsec:zak_ofdm_equiv}

We now demonstrate that the FD I/O relation proposed above is unitarily equivalent to conventional DD domain I/O relation with the pulsone basis per~\eqref{eq:zakotfs3}, $\boldsymbol{\phi}_{i}[n] = \mathbf{p}_{(k_0,l_0)}[n]$, $i = k_0 + l_0 M$, for which $\mathbf{s}[i] = \mathbf{X}[k_0,l_0]  \in \mathbb{C}^{MN\times 1}$. The corresponding DD domain vectorized system model is:
\begin{align}
    \label{eq:equiv1}
    \mathbf{r}_{{}_{\mathsf{DD}}} = \mathbf{H}_{{}_{\mathsf{DD}}} \mathbf{x}_{{}_{\mathsf{DD}}} + \mathbf{w}_{{}_{\mathsf{DD}}},
\end{align}
where $\mathbf{x}_{{}_{\mathsf{DD}}} = \mathsf{vec}(\mathbf{X})$. To relate the FD relation with the DD relation above, observe that the IDFZT in~\eqref{eq:zakofdm4} may be expressed in terms of $\mathbf{x}_{{}_{\mathsf{DD}}}$ as:
\begin{align}
    \label{eq:equiv2a}
    \mathbf{s} &= \mathbf{R} \mathbf{x}_{{}_{\mathsf{DD}}} \in \mathbb{C}^{MN\times 1}, \\
    \label{eq:equiv2b}
    \mathbf{R} &= \mathbf{K} (\mathbf{I}_{N} \otimes \mathbf{F}_{M}) \mathsf{diag}(\mathbf{q})  \in \mathbb{C}^{MN\times MN}, \\
    \label{eq:equiv2c}
    \mathbf{K} &= \sum_{i=1}^{N} \sum_{j=1}^{M} \mathbf{E}_{ij}^{\top} \otimes \mathbf{E}_{ij}  \in \{0, 1\}^{MN\times MN}, 
\end{align}
\begin{align}
    \label{eq:equiv2d}
    \mathbf{q} &= \begin{bmatrix}
        \vdots \\ \mathbf{q}_{l} \\ \vdots
    \end{bmatrix},~0 \leq l \leq (N-1) \in \mathbb{C}^{MN\times 1}, \\
    \label{eq:equiv2e}
    \mathbf{q}_{l}[k] &= e^{-j\frac{2\pi}{MN} lk}  \in \mathbb{C}^{M\times 1},~0 \leq k \leq (M-1),
\end{align}
where $\mathbf{E}_{ij} \in \mathbb{R}^{N\times M}$ with the $(i,j)$th entry equal to one and other entries zero, $\mathbf{F}_{M} \in \mathbb{C}^{M\times M}$ is the $M$-point DFT matrix. The matrix $\mathbf{R}$, therefore, represents the IDFZT operation.

\begin{lemma}
\label{lem:idfzt_unitary}
    The IDFZT matrix $\mathbf{R}$ is unitary.
\end{lemma}
\begin{IEEEproof}
To prove the Lemma, we must show that $\mathbf{R}^{\mathsf{H}}\mathbf{R} = \mathbf{R}\mathbf{R}^{\mathsf{H}} = \mathbf{I}_{MN}$. We evaluate:
\begin{align}
    \label{eq:r_unit_1}
    \mathbf{R}^{\mathsf{H}}\mathbf{R} &= (\mathsf{diag}(\mathbf{q}))^\mathsf{H}(\mathbf{I}_N\otimes \mathbf{F}_M)^\mathsf{H}\mathbf{K}^\mathsf{H}\mathbf{K}(\mathbf{I}_N\otimes\mathbf{F}_M)\mathsf{diag}(\mathbf{q}) \nonumber \\
    &\overset{(a)}{=} (\mathsf{diag}(\mathbf{q}))^\mathsf{H}(\mathbf{I}_N\otimes \mathbf{F}_M^\mathsf{H})(\mathbf{I}_N\otimes\mathbf{F}_M)\mathsf{diag}(\mathbf{q}) \nonumber \\
    &\overset{(b)}{=} (\mathsf{diag}(\mathbf{q}))^\mathsf{H}\mathsf{diag}(\mathbf{q}) \overset{(c)}{=} \mathbf{I}_{MN},
\end{align}
and similarly,
\begin{align}
    \label{eq:r_unit_2}
    \mathbf{R}\mathbf{R}^{\mathsf{H}} &= \mathbf{K}(\mathbf{I}_N\otimes\mathbf{F}_M)\mathsf{diag}(\mathbf{q})(\mathsf{diag}(\mathbf{q}))^\mathsf{H}(\mathbf{I}_N\otimes \mathbf{F}_M)^\mathsf{H}\mathbf{K}^\mathsf{H} \nonumber \\
    &\overset{(d)}{=} \mathbf{K}(\mathbf{I}_N\otimes\mathbf{F}_M)(\mathbf{I}_N\otimes \mathbf{F}_M^\mathsf{H})\mathbf{K}^\mathsf{H} \nonumber \\
    &\overset{(e)}{=} \mathbf{K}\mathbf{K}^\mathsf{H} \overset{(f)}{=} \mathbf{I}_{MN},
\end{align}
where steps $(a)$ and $(f)$ follow by observing that $\mathbf{K}$ is a permutation matrix and its inverse is its transpose, steps $(b)$ and $(e)$ follow from the orthonormality of the Fourier matrix, and steps $(c)$ and $(d)$ follow from the fact that the inverse of a complex phase is its conjugate.
\end{IEEEproof}


\begin{lemma}
    \label{lmm:zak_ofdm_unitary}
    The FD and DD domain I/O relations are unitarily equivalent.
\end{lemma}

\begin{IEEEproof}
From Lemma \ref{lem:idfzt_unitary},~\eqref{eq:equiv1} is equivalent to:
\begin{align}
    \label{eq:equiv3}
    \mathbf{r}_{{}_{\mathsf{DD}}} &= \mathbf{R}^{\mathsf{H}} \underbrace{\mathbf{R} \mathbf{H}_{{}_{\mathsf{DD}}} \mathbf{R}^{\mathsf{H}}}_{\mathbf{H}} \underbrace{\mathbf{R}\mathbf{x}_{{}_{\mathsf{DD}}}}_{\mathbf{s}} + \mathbf{w}_{{}_{\mathsf{DD}}}, \nonumber \\
    &= \mathbf{R}^{\mathsf{H}} \mathbf{H} \mathbf{s} + \mathbf{w}_{{}_{\mathsf{DD}}}.
\end{align}

Pre-multiplying~\eqref{eq:equiv3} by $\mathbf{R}$, we obtain~\eqref{eq:sys_model6}:
\begin{align}
    \label{eq:equiv4}
    \mathbf{r} = \mathbf{R} \mathbf{r}_{{}_{\mathsf{DD}}} = \mathbf{H} \mathbf{s} + \mathbf{w},
\end{align}
where $\mathbf{w} = \mathbf{R} \mathbf{w}_{{}_{\mathsf{DD}}}$. 
\end{IEEEproof}

\subsection{Embedding Information Symbols in FD}
\label{subsec:min_period_ofdm}

Instead of mounting the information symbols in the DD domain before conversion to the FD via the matrix $\mathbf{R}$ as in~\eqref{eq:equiv3}-\eqref{eq:equiv4}, an alternative is to mount information symbols directly in the FD, i.e.,
\begin{align}
    \label{eq:mp_ofdm1}
    \mathbf{r} = \mathbf{H} \mathbf{x}_{{}_{\mathsf{DD}}} + \mathbf{w},
\end{align}
where $\mathbf{H} = \mathbf{R} \mathbf{H}_{{}_{\mathsf{DD}}} \mathbf{R}^{\mathsf{H}}$ is the FD channel from~\eqref{eq:equiv3}.

Note that the input-output relation in~\eqref{eq:mp_ofdm1} corresponds to a type of FD modulation, where there are $MN$ carriers with carrier spacing $\Delta f = \nicefrac{B}{MN}$ in a single symbol of duration $T_{s} = \nicefrac{1}{\Delta f} = \nicefrac{MN}{B}$.




We have described three system models in \eqref{eq:sys_model6}, \eqref{eq:equiv4}, and \eqref{eq:mp_ofdm1}. It is natural to ask which one of them is the best in terms of performance. In the next Section, we try to answer this question.

\section{Selectivity Behavior of Different Methods}
\label{sec:optimal_basis}

\subsection{Recap of Proposed Methods}
\label{subsec:modulation_schemes}
We recap the methods described thus far.

In~\eqref{eq:sys_model6}, Zak-OTFS information symbols are mounted in the DD domain, which are transmitted through a DD channel. Equalization and symbol detection at the receiver also occurs in the DD domain.

In~\eqref{eq:equiv4}, Zak-OTFS information symbols are mounted in the DD domain, spreading the DD information symbols in the FD via the IDFZT matrix $\mathbf{R}$, which are then transmitted through a FD channel. Equalization at the receiver occurs in the FD, followed by de-spreading the FD equalized symbols through the conjugate transpose of the IDFZT matrix, $\mathbf{R}^{\mathsf{H}}$ to move from the FD to DD. Subsequent symbol detection occurs in the DD domain.

In~\eqref{eq:mp_ofdm1}, information symbols are mounted in the FD domain, which are transmitted through a FD channel. Equalization and symbol detection at the receiver also occurs in the FD. 

It is now natural to ask which method is better suited for time-varying fading channels. The fundamental difference between the methods is the choice of basis on which information symbols are mounted. In both FD and DD relations of Zak-OTFS, the basis used is the pulsone basis, while the DFT basis is used in~\eqref{eq:mp_ofdm1}. Therefore, the superiority of one method over another boils down to the choice of basis. We address this question next.

\subsection{Which Bases Are Non-Selective}?
\label{subsec:bases_non_fading}
As previously described, both FD and DD relations of Zak-OTFS mount information symbols on the pulsone basis (see \eqref{eq:zakotfs1a}), whereas~\eqref{eq:mp_ofdm1} mounts the information symbols on the IDFT basis (see \eqref{eq:zakofdm1}). In the following Lemma, we derive the condition for a basis to be non-selective. Subsequently, we show how the pulsone basis that enables both FD and DD relations of Zak-OTFS are \emph{non-selective} (i.e., the received energy is constant across all carriers). On the other hand, the choice of the IDFT basis makes~\eqref{eq:mp_ofdm1} and CP-OFDM \emph{selective} (i.e., the received energy is a function of carrier index). 

\begin{lemma}
    \label{lmm:non_fading}
    An orthonormal basis $\boldsymbol{\phi}_{i}[n]$ is non-selective if:
    \begin{align*}
        \boldsymbol{\phi}_{i}[n-k_2] \boldsymbol{\phi}_{i}^{*}[n-k_1] = \boldsymbol{\phi}_{j}[n-k_2] \boldsymbol{\phi}_{j}^{*}[n-k_1],
    \end{align*}
    for all $0 \leq i,j\leq (MN-1)$ and $0 \leq k_1,k_2 \leq (M-1)$, assuming the $MN$-periodized effective channel $\mathbf{h}[k,l]$ is supported within $0 \leq k \leq (M-1)$, $0 \leq l \leq (N-1)$.
\end{lemma}

\begin{IEEEproof}
If all carriers have equal received energy, the signal does not experience selectivity. Therefore, formally, the non-selectivity condition corresponds to:
\begin{align}
    \label{eq:non_fading1}
    (\mathbf{H}^{\mathsf{H}}\mathbf{H})[i,i] = (\mathbf{H}^{\mathsf{H}}\mathbf{H})[j,j],~\forall 0 \leq i,j \leq (MN-1),
\end{align}
that is, the value must be independent of the index ($i$ or $j$).

Substituting~\eqref{eq:sys_model5} in~\eqref{eq:non_fading1} and simplifying:
\begin{align}
    \label{eq:non_fading2}
    &(\mathbf{H}^{\mathsf{H}}\mathbf{H})[i,i] = \sum_{k_1,k_2} \sum_{l_1,l_2} \mathbf{h}^{*}[k_1,l_1] \mathbf{h}[k_2,l_2] e^{\frac{j2\pi}{MN}(k_1 l_1 - k_2 l_2)} \times \nonumber \\ &~~~\sum_{n_1,n_2} \bigg(\sum_{f=0}^{MN-1} \boldsymbol{\phi}_{f}[n_1] \boldsymbol{\phi}_{f}^{*}[n_2]\bigg) \boldsymbol{\phi}_{i}[n_2-k_2] \boldsymbol{\phi}_{i}^{*}[n_1-k_1] \times \nonumber \\ &~~~~~~~~~~~e^{\frac{j2\pi}{MN}(l_2 n_2 - l_1 n_1)}.
\end{align}

For an orthonormal basis, by definition, the inner summation over $f$ evaluates to $\delta[n_1-n_2]$. Therefore, $n_1 = n_2$ and hence:
\begin{align}
    \label{eq:non_fading3}
    (\mathbf{H}^{\mathsf{H}}\mathbf{H})[i,i] &= \sum_{k_1,k_2} \sum_{l_1,l_2} \mathbf{h}^{*}[k_1,l_1] \mathbf{h}[k_2,l_2] e^{\frac{j2\pi}{MN}(k_1 l_1 - k_2 l_2)} \times \nonumber \\ &~~~~\sum_{n} \boldsymbol{\phi}_{i}[n-k_2] \boldsymbol{\phi}_{i}^{*}[n-k_1] e^{\frac{j2\pi}{MN}(l_2 - l_1) n}.
\end{align}

Notice that the summation over $n$ is expansion along an orthonormal Fourier basis, and therefore $(\mathbf{H}^{\mathsf{H}}\mathbf{H})[i,i] = (\mathbf{H}^{\mathsf{H}}\mathbf{H})[j,j]$ implies that the summands must be equal.
Thus, the non-selectivity condition in~\eqref{eq:non_fading1} is satisfied for all orthonormal bases $\boldsymbol{\phi}_{i}[n]$ with the property:
\begin{align}
    \label{eq:non_fading4}
    \boldsymbol{\phi}_{i}[n-k_2] \boldsymbol{\phi}_{i}^{*}[n-k_1] = \boldsymbol{\phi}_{j}[n-k_2] \boldsymbol{\phi}_{j}^{*}[n-k_1],
\end{align}
for all $0 \leq i,j,k_1,k_2 \leq (MN-1)$. If the $MN$-periodized effective channel $\mathbf{h}[k,l]$ is supported within the fundamental period $0 \leq k \leq (M-1)$, $0 \leq l \leq (N-1)$, then~\eqref{eq:non_fading4} need only be satisfied for $0 \leq k_1,k_2 \leq (M-1)$.
\end{IEEEproof}

Via Lemma~\ref{lmm:non_fading}, we now quantify which bases are non-selective.

\subsubsection{Zak-OTFS in FD and DD}

For Zak-OTFS in the FD and DD domains we consider the pulsone basis, $\boldsymbol{\phi}_{i}[n] = \mathbf{p}_{(k_0,l_0)}[n]$, $i = k_0 + l_0 M$. From~\eqref{eq:zakotfs1a} we obtain:
\begin{align}
    \label{eq:non_fading5}
    &\boldsymbol{\phi}_{i}[n-k_2] \boldsymbol{\phi}_{i}^{*}[n-k_1] = \frac{1}{N} \sum_{d_1,d_2 \in \mathbb{Z}} e^{j\frac{2\pi}{N} (d_1-d_2) l_0} \times \nonumber \\ &~~~~\delta[n-k_0-k_1-d_1 M] \delta[n-k_0-k_2-d_2 M].
\end{align}

For $0 \leq k_1,k_2 \leq (M-1)$, the delta functions imply $k_1 = k_2$ and $0 \leq d_1 = d_2 \leq (N-1)$. Therefore,
\begin{align}
    \label{eq:non_fading6}
    \boldsymbol{\phi}_{i}[n-k_2] \boldsymbol{\phi}_{i}^{*}[n-k_1] &= \delta[k_1-k_2],
\end{align}
which is independent of $0 \leq i \leq (MN-1)$, i.e., Zak-OTFS \emph{does not exhibit selectivity}.

\subsubsection{FD I/O Relation in~\eqref{eq:mp_ofdm1}}

In~\eqref{eq:mp_ofdm1}, we consider the IDFT basis in~\eqref{eq:zakofdm1}. We have:
\begin{align}
    \label{eq:non_fading7}
    \boldsymbol{\phi}_{i}[n-k_2] \boldsymbol{\phi}_{i}^{*}[n-k_1] &= \frac{1}{MN}  e^{j\frac{2\pi}{MN} i (k_1 - k_2)},
\end{align}
which depends on $0 \leq i \leq (MN-1)$, i.e., it \emph{exhibits selectivity}.


Similar analysis is possible also for CP-OFDM, but is not described due to space constraints.

\begin{figure}
    \centering
    \includegraphics[width=0.9\linewidth]{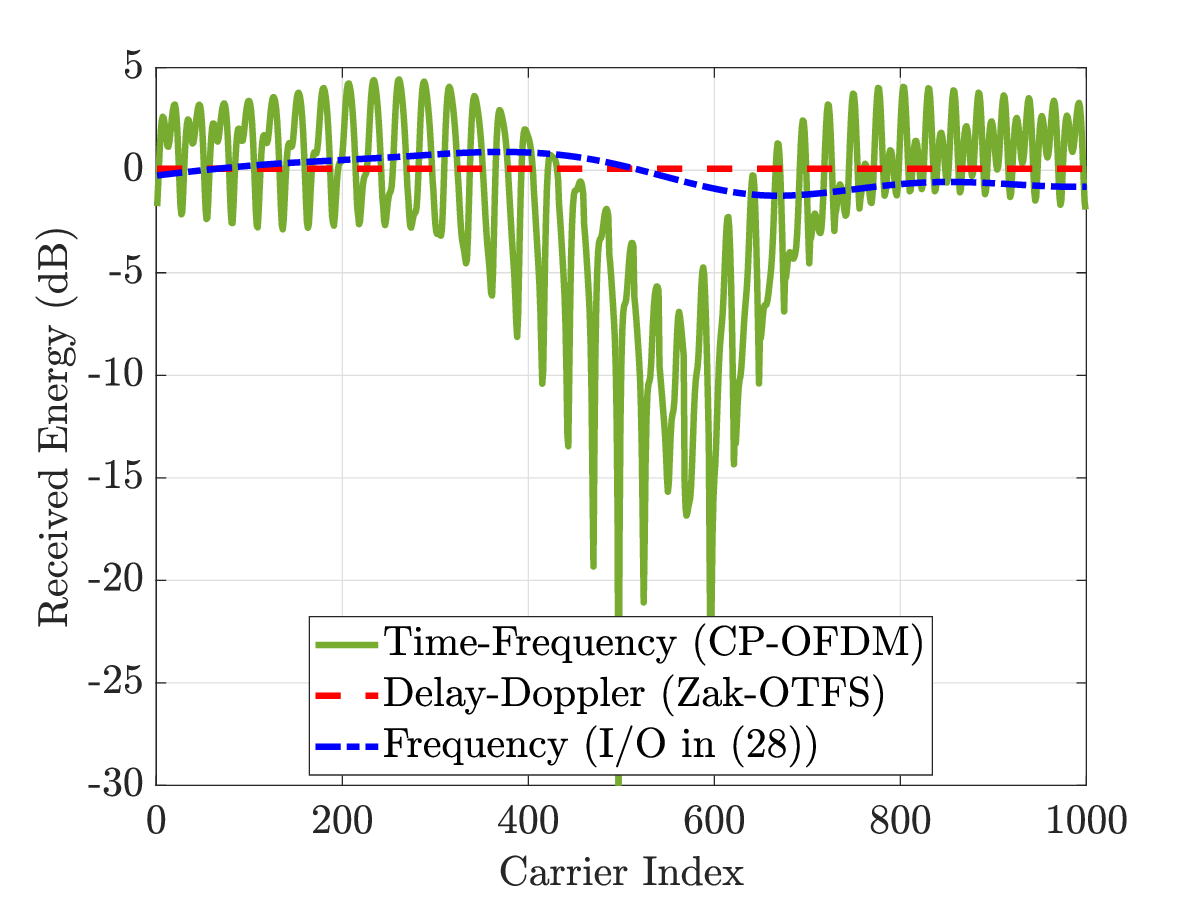}
    \caption{Received energy per carrier showing how Zak-OTFS is non-selective, whereas the FD I/O relation in~\eqref{eq:mp_ofdm1} and CP-OFDM fade.} 
    \label{fig:column_norm_comp}
\end{figure}

Fig.~\ref{fig:column_norm_comp} shows the received energy per carrier (the squared column norm of the channel matrix) for Zak-OTFS, I/O relation in~\eqref{eq:mp_ofdm1}, and CP-OFDM, for a six-path Vehicular-A channel (Table~\ref{tab:veh_a}) with $2.51 \mu$s delay spread and $815$ Hz Doppler spread. Consistent with the above discussion, we observe that the received energy per carrier remains constant for Zak-OTFS. On the other hand, the received energy per carrier varies for the I/O relation in~\eqref{eq:mp_ofdm1} and CP-OFDM due to their fading characteristics. Note that the received energy is plotted in the dB scale. Therefore, although the variation for the I/O relation in~\eqref{eq:mp_ofdm1} seems small (in dB), the variance of the received energy is significant in linear scale. Furthermore, the variation for CP-OFDM is larger than that of~\eqref{eq:mp_ofdm1}, suggesting greater resilience to Doppler of the latter. Later in Sec.~\ref{sec:results} we present results consistent with the findings of Fig.~\ref{fig:column_norm_comp}.

\textit{Remark:} Why is the basis being non-selective important? For a modulation scheme with non-selective basis, the received energy is constant irrespective of the carrier index (Zak-OTFS in Fig.~\ref{fig:column_norm_comp}). This largely simplifies channel coding. On the other hand, if the basis is selective (CP-OFDM in Fig.~\ref{fig:column_norm_comp}), channel coding needs to be carried out across subcarriers since the received energy is a function of the subcarrier index. Such coding helps the symbols experience channels from different subcarriers and improves the SNR at the receiver. The encoding and decoding complexity of such codes is typically higher (for example, LDPC codes in OFDM).

In the next Section, we describe the proposed low complexity equalization and detection scheme that makes use of the FD perspective of Zak-OTFS.




\section{Low Complexity FD Equalization \& Detection}
\label{sec:low_compl_eq}

\begin{figure}
    \centering
    \includegraphics[width=0.9\linewidth]{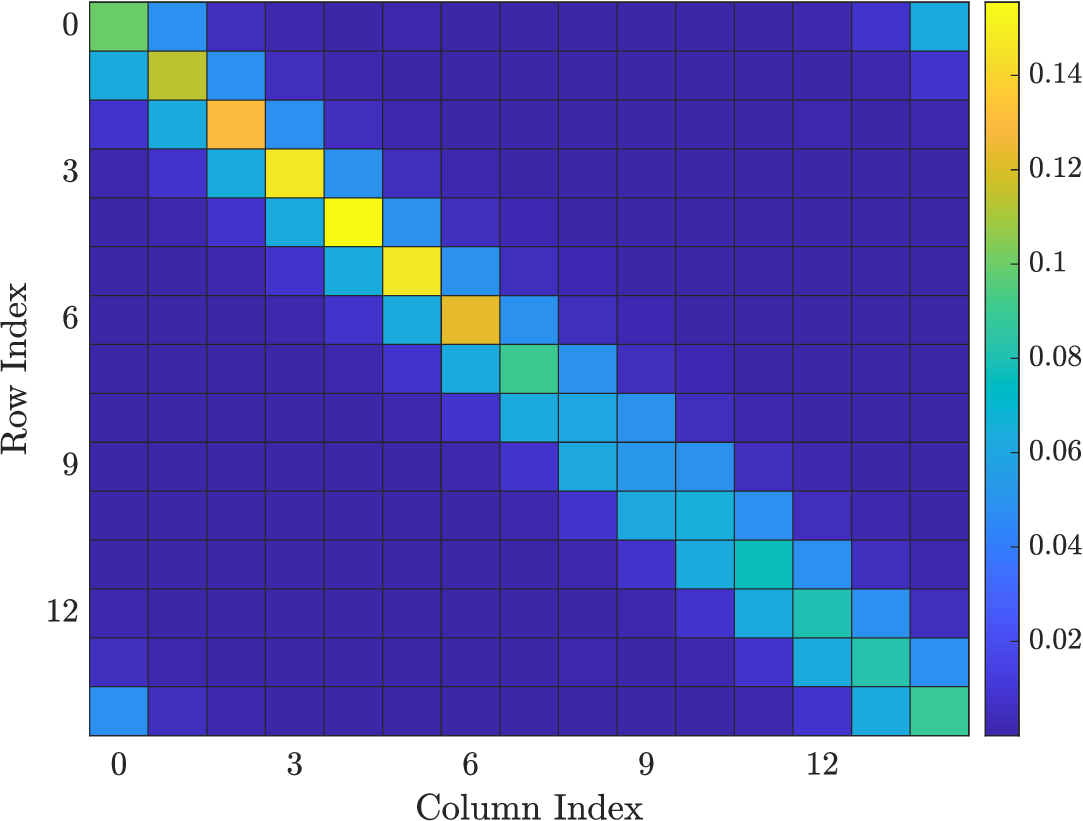}
    \caption{Modulo banded structure in the FD channel matrix $\mathbf{H}$ for $M=3, N=5, \nu_p = 30$ kHz, and the Veh-A channel (Table~\ref{tab:veh_a}).}
    \label{fig:mod_banded}
\end{figure}

To develop low complexity equalization \& detection algorithms, we consider the FD channel matrix, $\mathbf{H}$, in \eqref{eq:equiv4}. The entries of $\mathbf{H}$ are given by \eqref{eq:zak_ofdm6}:
\begin{align*}
    \mathbf{H}[f,i] &= \sum_{\bar{k}=0}^{MN-1} \mathbf{h}[\bar{k},(f-i) \bmod MN] e^{-\frac{j2\pi}{MN}f\bar{k}}.
\end{align*}
Let $\mathcal{S}$ denote the support of the channel in $\mathbf{h}$ in an $MN \times MN$ region. For a given row index $f$, and $i = 0, 1, \cdots, MN-1$, notice that as long as $(f-i)\bmod MN$ is within $\mathcal{S}$, the corresponding entry within $\mathbf{H}$ is non-zero. Therefore the matrix $\mathbf{H}$ has a modulo-banded structure. To demonstrate this, the absolute value of $\mathbf{H}$ is plotted in Fig. \ref{fig:mod_banded} for $M=3, N = 5$. The sub- and super-diagonals fold back into the right-top and left-bottom corners, respectively, of the matrix. The non-zero values at the corners of the matrix disallow low complexity equalization. In the following Subsection, we describe a method to effectively handle these values and develop a low-complexity equalization algorithm.

\textit{Remark:} The response of the channel to a tone (or pulse in the FD) is fixed irrespective of the modulation scheme considered. Estimating this response is possible as long as there is no ICI due to Doppler. However, under doubly-dispersive channels, the effective channel matrix is no longer diagonal (see Fig.~\ref{fig:mod_banded}). A pilot in the FD is only able to estimate the diagonal entries in the matrix, which do not capture the interaction between subcarriers due to Doppler (because of ICI) and leads to poor equalization and data detection performance. Here, we capture the I/O relation in the DD domain and move it to FD, so that we have the estimate of not just the diagonal values, but also off-diagonal values in the FD channel matrix.

\subsection{Reduction to Banded Structure}
\label{subsec:red_to_truly_banded_struct}

To reduce the modulo-banded structure of $\mathbf{H}$ to an actual banded structure, we propose to zero out the initial and final few entries in $\mathbf{s}$. However, directly nulling out the first and last few entries of $\mathbf{s}$ is not feasible due to the relation $\mathbf{s} = \mathbf{R}\mathbf{x}_{{}_{\mathsf{DD}}}$, since the available degrees-of-freedom is only over $\mathbf{x}_{{}_{\mathsf{DD}}}$. To overcome this challenge, we propose the following approach.

Let $b$ denote the width of the band on either side of the diagonal, referred to as spread width. For example, in Fig. \ref{fig:mod_banded}, $b=2$, which denotes the number of significant non-zero entries on either side of the diagonal. Notice that the significant values in the corner are also limited to the $b\times b$ square. We construct a matrix $\mathbf{R}'$ as:
\begin{align}
    \label{eq:low_comp_eq1}
    \mathbf{R}' = \begin{bmatrix}
        \mathbf{R}_{0:b-1} \\
        \mathbf{R}_{MN-b:MN-1}
    \end{bmatrix}\in \mathbb{C}^{2b\times MN},
\end{align}
where $\mathbf{R}_{0:b-1}$ denotes the first $b$ rows in $\mathbf{R}$ and $\mathbf{R}_{MN-b:MN-1}$ denotes the last $b$ rows in $\mathbf{R}$. The rank of $\mathbf{R}'$ is $2b$ and the dimension of the null space is $(MN-2b)$ (since the number of colums is $MN$). 

Let $\mathbf{N} = \mathsf{null}(\mathbf{R}') \in \mathbb{C}^{MN\times MN-2b}$ denote the null space of $\mathbf{R}'$. We transmit the FD vector:
\begin{align}
    \label{eq:low_comp_eq2}
    \mathbf{s}' = \mathbf{R}\mathbf{N}\mathbf{x}'_{{}_{\mathsf{DD}}} \in \mathbb{C}^{MN\times 1},
\end{align}
where $\mathbf{x}'_{{}_{\mathsf{DD}}}$ is an $(MN-2b) \times 1$ vector\footnote{Although we lose spectral efficiency by using this approach, the spectral efficiency loss is still very small compared to cyclic prefix in OFDM~\cite{khan2025waveform} and guard symbols in an embedded pilot frame.\label{footnote:se_loss}} of information symbols. With this approach, the system model becomes:
\begin{align}
    \label{eq:low_comp_eq3}
    \mathbf{r}' = \mathbf{H}\mathbf{s}' + \mathbf{w}.
\end{align}

Our approach ensures that the first $b$ and last $b$ entries in $\mathbf{s}'$ are zeros, and only the banded portion of the channel matrix $\mathbf{H}$ impacts the transmitted symbols.

\subsection{Low-Complexity MMSE Equalization}
\label{subsec:low_compl_eq}

To detect the information symbols, $\mathbf{r}'$ in~\eqref{eq:low_comp_eq3} is equalized using $\mathbf{H}$. The banded structure of $\mathbf{H}$ 
enables the development of
low-complexity equalization routines.

We consider the linear minimum mean square error (LMMSE) equalizer $\mathbf{T} \in \mathbb{C}^{MN\times MN}$ given by:
\begin{align}
    \label{eq:low_comp_eq5}
    \mathbf{T} = \mathbf{H}^{\mathsf{H}}(\mathbf{H}\mathbf{H}^\mathsf{H} + \mathbf{R}_{n})^{-1} = (\mathbf{I}_{MN}+\mathbf{H}^{\mathsf{H}}\mathbf{R}_n^{-1}\mathbf{H})^{-1}\mathbf{H}^{\mathsf{H}}\mathbf{R}_n^{-1},
\end{align}
where $\mathbf{R}_{n} \in \mathbb{C}^{MN\times MN}$ is the covariance matrix of noise, and $\mathbf{H}$ is the channel matrix. Inverting the matrix $(\mathbf{I}_{MN}+\mathbf{H}^{\mathsf{H}}\mathbf{R}_n^{-1}\mathbf{H})$ in the absence of any structure incurs complexity $\mathcal{O}(M^3N^3)$.

However, since $\mathbf{H}$ is banded, $\mathbf{T}$ is also banded. For banded matrices, low-complexity LMMSE matrix computation is feasible \cite{Huang2009_bandedinversion, Philipp1991_bandedinversion}. In the following, we describe one feasible algorithm based on the conjugate gradient method (CGM)~\cite{liu2020energy}.

Consider the LMMSE equalization operation:
\begin{align}
    \label{eq:low_comp_eq6}
    \tilde{\mathbf{s}} = \mathbf{T}\mathbf{r}' \implies \underbrace{(\mathbf{I}_{MN}+\mathbf{H}^{\mathsf{H}}\mathbf{R}_n^{-1}\mathbf{H})}_{\mathbf{Q}}\tilde{\mathbf{s}} = \underbrace{\mathbf{H}^{\mathsf{H}}\mathbf{R}_n^{-1}\mathbf{r}'}_{\mathbf{t}}.
\end{align}
The linear system is of the form $\mathbf{Q}\tilde{\mathbf{s}} = \mathbf{t}$, where $\mathbf{Q}$ is Hermitian positive definite\footnote{A matrix $\mathbf{A} \succ 0$ if $\mathbf{x^{\mathsf{H}}Ax}>0$ for any non-zero $\mathbf{x}$. $\mathbf{x}^{\mathsf{H}}\mathbf{Q}\mathbf{x} = \mathbf{x}^{\mathsf{H}}\mathbf{x} + (\mathbf{Hx})^{\mathsf{H}}\mathbf{R}_n^{-1}\mathbf{Hx} > 0$ since $\mathbf{R}_n \succ 0$ (covariance matrix of noise is Hermitian positive definite).}\label{footnote:psd} ($\mathbf{Q} \succ 0$).
The conjugate gradient method
solves $\mathbf{Q}\tilde{\mathbf{s}} = \mathbf{t}$ by minimizing the quadratic energy:
\begin{align}
\label{eq:low_comp_eq7}
    \mathcal{J}(\tilde{\mathbf{s}}) = \frac{1}{2} \tilde{\mathbf{s}}^\mathsf{H} \mathbf{Q} \tilde{\mathbf{s}} - \mathbf{t}^\mathsf{H} \tilde{\mathbf{s}},    
\end{align}
since this has a unique minimum at:
\begin{align}
    \label{eq:low_comp_eq8}
    \nabla \mathcal{J}(\tilde{\mathbf{s}}) = \mathbf{Q}\tilde{\mathbf{s}} - \mathbf{t} = 0 \implies \tilde{\mathbf{s}} = \mathbf{Q}^{-1}\mathbf{t},
\end{align}
which is the desired LMMSE output.
This function in~\eqref{eq:low_comp_eq7} is strictly convex when $\mathbf{Q} \succ 0$, and has a unique minimizer at $\tilde{\mathbf{s}} = \mathbf{Q}^{-1}\mathbf{t}$. 
The conjugate gradient algorithm
minimizes this function iteratively by moving along conjugate directions $\mathbf{p}^{(i)}$, which are chosen so that $\mathbf{p}^{(i)\mathsf{H}} \mathbf{Q} \mathbf{p}^{(j)} = 0$ for $i \ne j$. These directions are ``$\mathbf{Q}$-orthogonal'', meaning no search direction undoes the progress of a previous one. Next, the solution is updated via:
\begin{align}
    \label{eq:low_comp_eq9}
    \tilde{\mathbf{s}}^{(i+1)} = \tilde{\mathbf{s}}^{(i)} + \alpha \mathbf{p}^{(i)},
\end{align}
where $\alpha$ is chosen to minimize $\mathcal{J}(\tilde{\mathbf{s}})$ along the direction $\mathbf{p}^{(i)}$. Finally, the residual is updated via:
\begin{align}
    \label{eq:low_comp_eq10}
    \mathbf{c}^{(i)} = \mathbf{t} - \mathbf{Q}\tilde{\mathbf{s}}^{(i)} ,
\end{align}
which becomes orthogonal to all previous residuals:
\begin{align}
    \label{eq:low_comp_eq11}
    \mathbf{c}^{(i)\mathsf{H}} \mathbf{c}^{(j)} = 0 \quad \text{for } i \ne j.
\end{align}

The steps of the iterative conjugate gradient method are listed in Algorithm~\ref{alg:conj_grad}.
 
\begin{algorithm}
    \caption{Iterative conjugate gradient algorithm}
    \label{alg:conj_grad}
    \begin{algorithmic}[1]
        \STATE \textbf{Inputs:} Channel matrix $\mathbf{H}$, received vector $\mathbf{r}$, covariance matrix of noise $\mathbf{R}_n$, number of delay bins $M$, number of Doppler bins $N$, tolerance $\epsilon$, maximum number of iterations $\max\_\mathrm{iter}$
        \STATE \textbf{Initialize:} $\breve{\mathbf{H}} = \mathbf{H}^{\mathsf{H}}\mathbf{H}$, $\mathbf{b} = \mathbf{H}^{\mathsf{H}}\mathbf{r}$, $\tilde{\mathbf{s}}^{(0)} = \mathbf{0}_{MN\times 1}$, $\mathbf{c}^{(0)} = \mathbf{b} - \breve{\mathbf{H}}\tilde{\mathbf{s}}^{(0)}-\mathbf{R}_n\tilde{\mathbf{s}}^{(0)}$, $\mathbf{p}^{(0)} = \mathbf{c}^{(0)}$, $c^{(0)}_{\mathsf{norm}} = \Vert\mathbf{c}\Vert_2^2$
        \FOR {$i=1:\max\_\mathrm{iter}$}
            \STATE $\mathbf{a}_{\mathbf{p}} = \breve{\mathbf{H}}{\mathbf{p}^{(i-1)}}+\mathbf{R}_n\mathbf{p}^{(i-1)}$
            \STATE $\alpha = \frac{c^{(i-1)}_{\mathsf{norm}}}{\mathbf{p}^{\mathsf{H}}\mathbf{a}_{\mathbf{p}}}$
            \STATE $\tilde{\mathbf{s}}^{(i)} = \tilde{\mathbf{s}}^{(i-1)} + \alpha \mathbf{p}^{(i-1)}$
            \STATE $\mathbf{c}^{(i)} = \mathbf{c}^{(i-1)} - \alpha\mathbf{a}_{\mathbf{p}}$
            \STATE $c^{(i)}_{\mathsf{norm}} = \Vert\mathbf{c}^{(i)}\Vert_2^2$
            \IF{$c^{(i)}_{\mathsf{norm}} < \epsilon^2$}
                \STATE break
            \ENDIF
            \STATE $\mathbf{p}^{(i)} = \mathbf{c}^{(i)} + \frac{c^{(i)}_{\mathsf{norm}}}{c^{(i-1)}_{\mathsf{norm}}}\mathbf{p}^{(i-1)}$
            \STATE $c^{(i-1)}_{\mathsf{norm}} = c^{(i)}_{\mathsf{norm}}$
        \ENDFOR
        \RETURN $\tilde{\mathbf{s}}^{(i)}$
    \end{algorithmic}
\end{algorithm}

\subsubsection{Complexity}
\label{subsubsec:complexity}
\begin{table}
    \centering
    \caption{Complexity per iteration of the iterative conjugate gradient algorithm.}
    \begin{tabular}{|c|c|}
         \hline 
         Operation & Complexity \\
         \hline
         $\mathbf{a}_{\mathbf{p}} = \breve{\mathbf{H}}{\mathbf{p}^{(i-1)}}+\mathbf{R}_n\mathbf{p}^{(i-1)}$ & $\mathcal{O}(bMN)$ \\ 
         $\alpha = \frac{c^{(i-1)}_{\mathsf{norm}}}{\mathbf{p}^{\mathsf{H}}\mathbf{a}_{\mathbf{p}}}$ & $\mathcal{O}(MN)$\\
         $\tilde{\mathbf{s}}^{(i)} = \tilde{\mathbf{s}}^{(i-1)} + \alpha \mathbf{p}^{(i-1)}$ & $\mathcal{O}(MN)$\\
         $\mathbf{c}^{(i)} = \mathbf{c}^{(i-1)} - \alpha\mathbf{a}_{\mathbf{p}}$ & $\mathcal{O}(MN)$\\
         $c^{(i)}_{\mathsf{norm}} = \Vert\mathbf{c}^{(i)}\Vert_2^2$ & $\mathcal{O}(MN)$\\
         $\mathbf{p}^{(i)} = \mathbf{c}^{(i)} + \frac{c^{(i)}_{\mathsf{norm}}}{c^{(i-1)}_{\mathsf{norm}}}\mathbf{p}^{(i-1)}$ & $\mathcal{O}(MN)$\\
         $c^{(i-1)}_{\mathsf{norm}} = c^{(i)}_{\mathsf{norm}}$ & $1$\\
         \hline
    \end{tabular}
    
    \label{tab:cg_complexity}
\end{table}
The complexity involved in each iteration of the conjugate gradient algorithm in Algorithm~\ref{alg:conj_grad} is presented in Table~\ref{tab:cg_complexity}. The complexity is dominated by the Step 4 in Algorithm~\ref{alg:conj_grad} which is of the order of $bMN$. The overall complexity of the iterative conjugate gradient algorithm is therefore $\mathcal{O}(kbMN)$, where $k$ is the number of iterations before convergence. Setting $k = \max\_\mathrm{iter}$ gives the worst case complexity of Algorithm~\ref{alg:conj_grad}. On the other hand, the DD domain Zak-OTFS channel matrix does not have a banded structure and this incurs complexity $\mathcal{O}(kMN^2)$ \cite{liu2020energy}.

\subsection{Information Symbol Detection}
\label{subsec:sym_detecn}

Let $\tilde{\mathbf{s}}$ denote the vector post-equalization, i.e., $\tilde{\mathbf{s}} = \mathbf{T} \mathbf{r}'$. Information symbols are detected via minimum distance decoding:
\begin{align}
    \label{eq:low_comp_eq4}
    \widehat{\mathbf{x}'}_{{}_{\mathsf{DD}}}[q] = \underset{a \in \mathcal{A}}{\arg \min}\vert(\mathbf{N}^{\mathsf{H}}\mathbf{R}^\mathsf{H}\tilde{\mathbf{s}})[q] - a\vert,
\end{align}
where $q = 0, 1, \cdots, (MN-2b-1)$, and $\mathcal{A}$ is the constellation (e.g., $4$-QAM) from which information symbols are drawn.

The discussion presented up to this point assumes perfect channel knowledge. However, this is not true in practice. In the next Section we present a channel estimation scheme to estimate the FD channel matrix.

\section{Practical Channel Estimation}
\label{sec:channel_est}
Channel estimation is a crucial aspect of any practical system. In this paper, we consider the spread pilot~\cite{Aug2024paper, preamblepaper} for channel estimation.

\begin{figure*}
    \centering
    \includegraphics[width=0.7\linewidth]{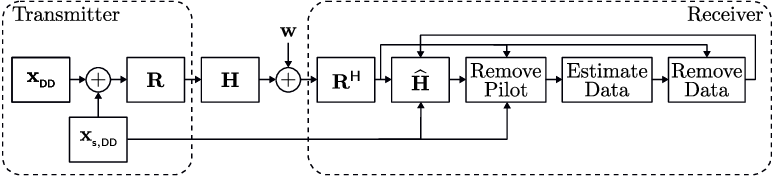}
    \caption{Block diagram representing the flow of the channel estimation algorithm.}
    \label{fig:ch_est_block_dia}
\end{figure*}

The block diagram of the channel estimation algorithm is presented in Fig. \ref{fig:ch_est_block_dia}. At the transmitter, along with the information symbols\footnote{For ease of exposition, in this description, we ignore the step of mounting the information symbols on the null space of $\mathbf{R}'$ presented in Section \ref{subsec:red_to_truly_banded_struct}. However, for all the simulations, we do perform this step as $\sqrt{e_p}\mathbf{x}_{{}_{\mathsf{s, DD}}}+\sqrt{e_d}\mathbf{N}\mathbf{x}'_{{}_{\mathsf{DD}}}$.} in $\mathbf{x}_{{}_{\mathsf{DD}}} \in \mathbb{C}^{MN \times 1}$ the spread pilot $\mathbf{x}_{{}_{\mathsf{s, DD}}} \in \mathbb{C}^{MN \times 1}$ is added at each location, i.e., the spread pilot is superimposed with the information symbols. The DD domain vectors are then transformed to the FD using the IDFZT matrix $\mathbf{R}$. The received vector is converted back to the DD domain using the inverse of the IDFZT matrix, $\mathbf{R}^{\mathsf{H}}$. The effective system model is:
\begin{align}
    \label{eq:ch_est_1}
    \mathbf{y}_{{}_{\mathsf{DD}}} = \mathbf{R}^{\mathsf{H}}\mathbf{H}\mathbf{R}(\sqrt{e_d}\mathbf{x}_{{}_{\mathsf{DD}}}+\sqrt{e_p}\mathbf{x}_{{}_{\mathsf{s, DD}}}), 
\end{align}
where $e_p$ and $e_d$ denote the energy in the pilot and data symbols, respectively. At the receiver, the model-free estimate of the effective channel $\widehat{\mathbf{h}}_\mathsf{eff}$ (in \eqref{eq:sys_model1}) is obtained through the cross-ambiguity function between $\mathbf{y}_{{}_{\mathsf{DD}}}$ and $\mathbf{x}_{{}_{\mathsf{s, DD}}}$ \cite{Aug2024paper}, \cite{preamblepaper}:
\begin{align}
    \label{eq:ch_est_2}
    \widehat{\mathbf{h}}_\mathsf{eff}[k, l] = \sum_{k'=0}^{M-1}\sum_{l'=0}^{N-1}&\mathbf{y}_{{}_{\mathsf{DD}}}[k', l']\mathbf{x}^*_{{}_{\mathsf{s, DD}}}[k'-k, l'-l]\times \nonumber \\
    &e^{-j\frac{2\pi}{MN}l'(k'-k)}.
\end{align}
Next the estimated channel matrix $\widehat{\mathbf{H}}_{{}_{\mathsf{DD}}}$ (see \eqref{eq:equiv1}) is constructed using the estimated $\widehat{\mathbf{h}}_\mathsf{eff}$. 
An estimate of the channel matrix in \eqref{eq:equiv4}, $\widehat{\mathbf{H}}$ is obtained as (from \eqref{eq:equiv3}):
\begin{align}
    \label{eq:ch_est_3}
    \widehat{\mathbf{H}} = \mathbf{R}\widehat{\mathbf{H}}_{{}_{\mathsf{DD}}}\mathbf{R}^{\mathsf{H}}.
\end{align}
For data detection, the pilot is removed from the received frame using the estimated channel as:
\begin{align}
    \label{eq:ch_est_4}
    \mathbf{y}_{{}_{\mathsf{DD, data}}} = \mathbf{y}_{{}_{\mathsf{DD}}} - \sqrt{e_p}\widehat{\mathbf{H}}_{{}_{\mathsf{DD}}}\mathbf{x}_{{}_{\mathsf{s, DD}}}.
\end{align}
The data-only frame is then converted to FD using the IDFZT matrix and low-complexity  equalization and detection is performed as described in Section \ref{sec:low_compl_eq} to obtain the detected information symbols $\widehat{\mathbf{x}}_{{}_{\mathsf{DD}}}$.

\begin{table}[!t]
    \centering
    \caption{Power-delay profile of Veh-A channel model}
    \begin{tabular}{|c|c|c|c|c|c|c|}
         \hline
         Path index $i$ & 1 & 2 & 3 & 4 & 5 & 6 \\
         \hline
         Delay $\tau_i (\mu s)$ & 0 & 0.31 & 0.71 & 1.09 & 1.73 & 2.51 \\
         \hline
         Relative power (dB) & 0 & -1 & -9 & -10 & -15 & -20 \\
         \hline
    \end{tabular}
    \label{tab:veh_a}
\end{table}

\subsection{Turbo Iterations}
\label{subsec:turbo_iter}
Note that, in~\eqref{eq:ch_est_4}, 
the estimated channel matrix is used to remove the pilot.
However, since the estimate is obtained in the presence of interference from data symbols, the estimate is noisy and leads to inefficient pilot removal. This affects the detection performance. To improve the performance, we perform turbo iterations which alternate between data detection and channel estimation as described below.

The estimated data is removed from the received frame as:
\begin{align}
    \label{eq:ch_est_5}
    \mathbf{y}_{{}_{\mathsf{DD, pil}}} = \mathbf{y}_{{}_{\mathsf{DD}}} - \sqrt{e_d}\widehat{\mathbf{H}}_{{}_{\mathsf{DD}}}\widehat{\mathbf{x}}_{{}_{\mathsf{DD}}},
\end{align}
to obtain the pilot-only frame. Channel estimation in~\eqref{eq:ch_est_2} is performed by replacing $\mathbf{y}_{{}_{\mathsf{DD}}}$ with $\mathbf{y}_{{}_{\mathsf{DD, pil}}}$ which leads to a better estimate on account of reduced interference from data symbols. The estimated channel matrices are reconstructed as shown in \eqref{eq:ch_est_3}. The data only frame is obtained through \eqref{eq:ch_est_4} and data equalization and detection is performed on this frame. The detected data is used in \eqref{eq:ch_est_5} and this procedure is repeated a fixed number of times. Better channel estimation accuracy aids detection accuracy which in turn aids channel estimation accuracy. Therefore the performance of the estimation algorithm improves with turbo iterations.

In the following, we present results that evaluate the performance of the proposed scheme under perfect and imperfect channel knowledge and with various pulse shapes.

\begin{figure}
    \centering
    \includegraphics[width=0.99\linewidth]{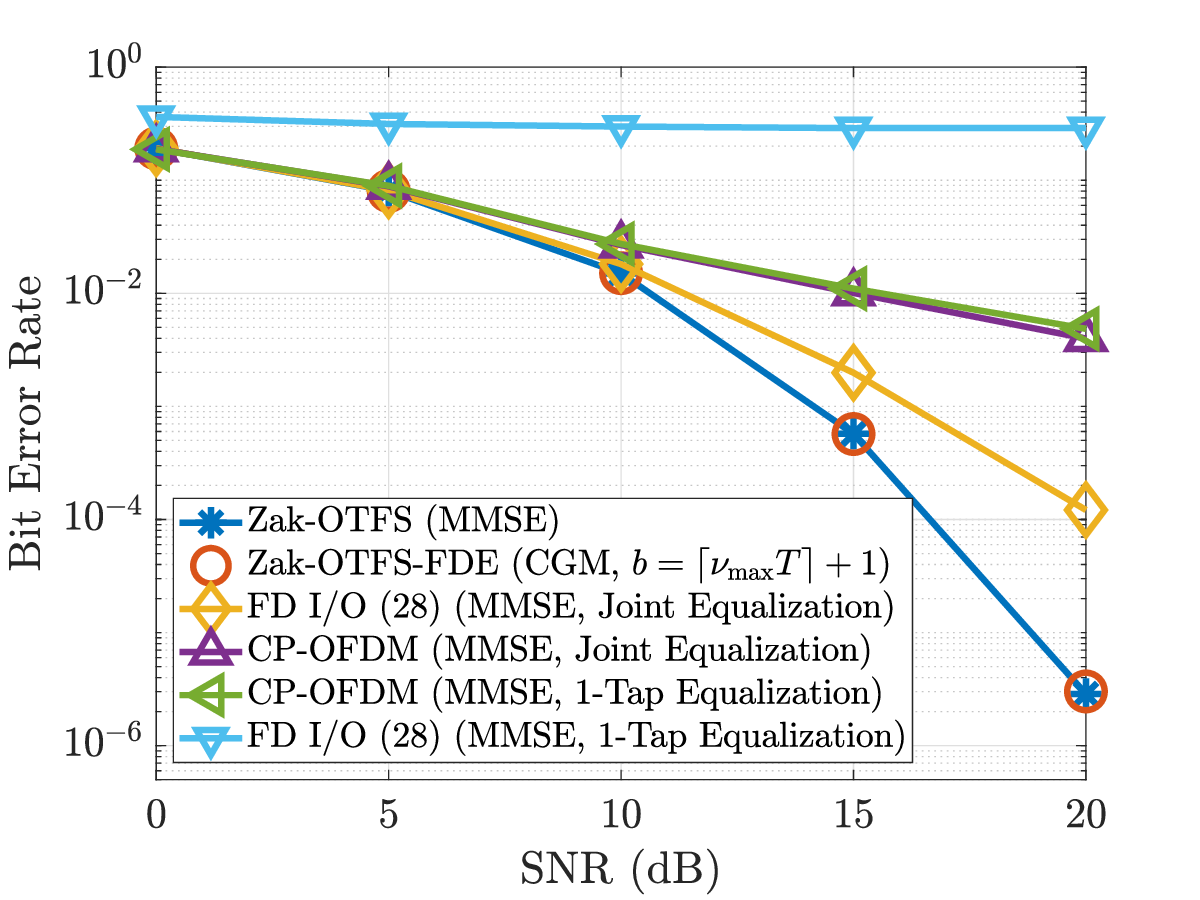}
    \caption{Comparison of BER of different modulation schemes with perfect channel knowledge at high Doppler values. FD equalization (FDE) of Zak-OTFS via CGM algorithm with $k=250, \epsilon=10^{-6}, b=\lceil\nu_{\max}T\rceil+1$. Simulation parameters: $M = 31$, $N = 37$, $\nu_{p} = 30$ kHz, Veh-A channel with $\nu_{\max} = 815$ Hz, RRC pulse shaping filter with $\beta_\tau = \beta_\nu = 0.6$.}
    \label{fig:perf_csi_vs_algo_M31_N37_highdop}
\end{figure}

\begin{figure}
    \centering
    \includegraphics[width=0.99\linewidth]{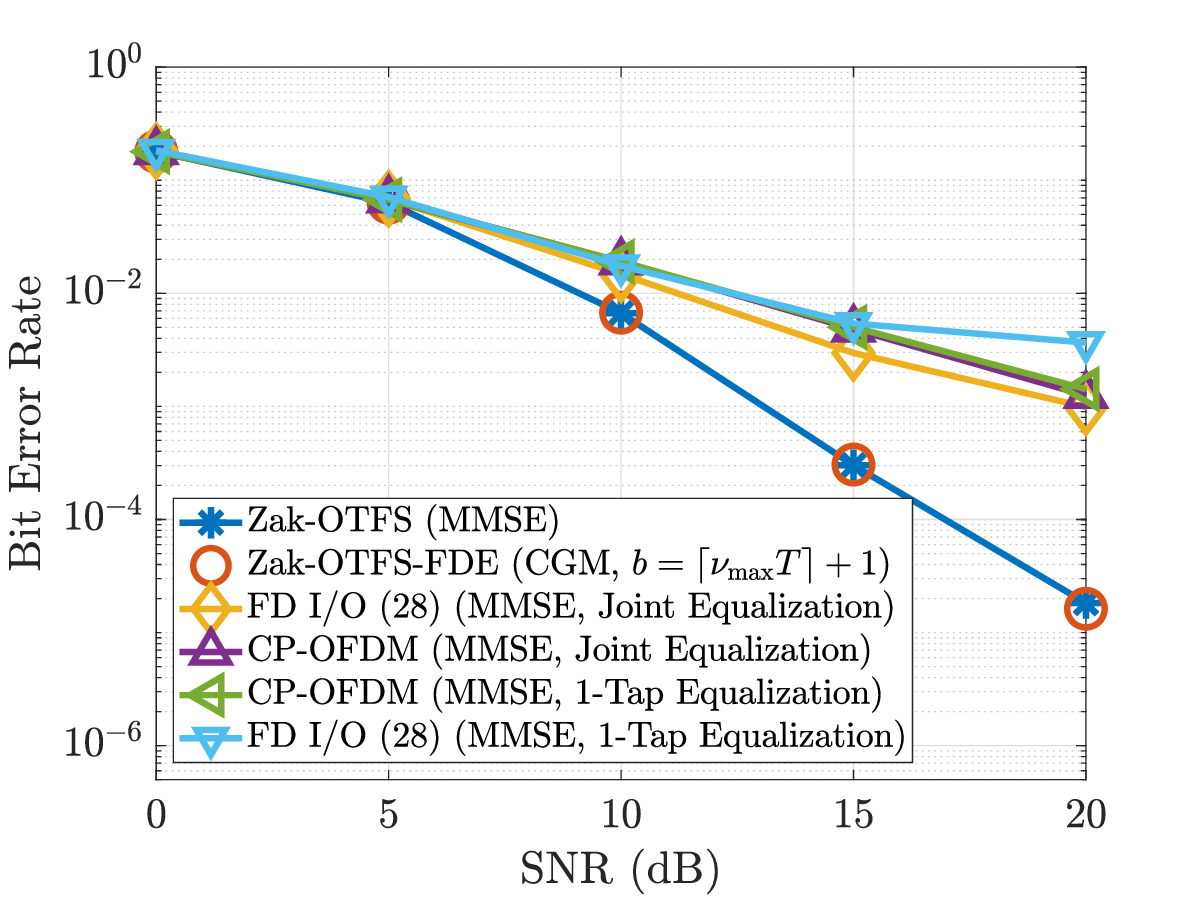}
    \caption{Comparison of BER of different modulation schemes with perfect channel knowledge at low Doppler values. FD equalization (FDE) of Zak-OTFS via CGM algorithm with $k=250, \epsilon=10^{-6}, b=\lceil\nu_{\max}T\rceil+1$. Simulation parameters: $M = 31$, $N = 37$, $\nu_{p} = 30$ kHz, Veh-A channel with $\nu_{\max} = 81.5$ Hz, RRC pulse shaping filter with $\beta_\tau = \beta_\nu = 0.6$.}
    \label{fig:perf_csi_vs_algo_M31_N37_lowdop}
\end{figure}

\begin{figure*}
    \subfloat[{NMSE}]{\includegraphics[width=0.45\linewidth]{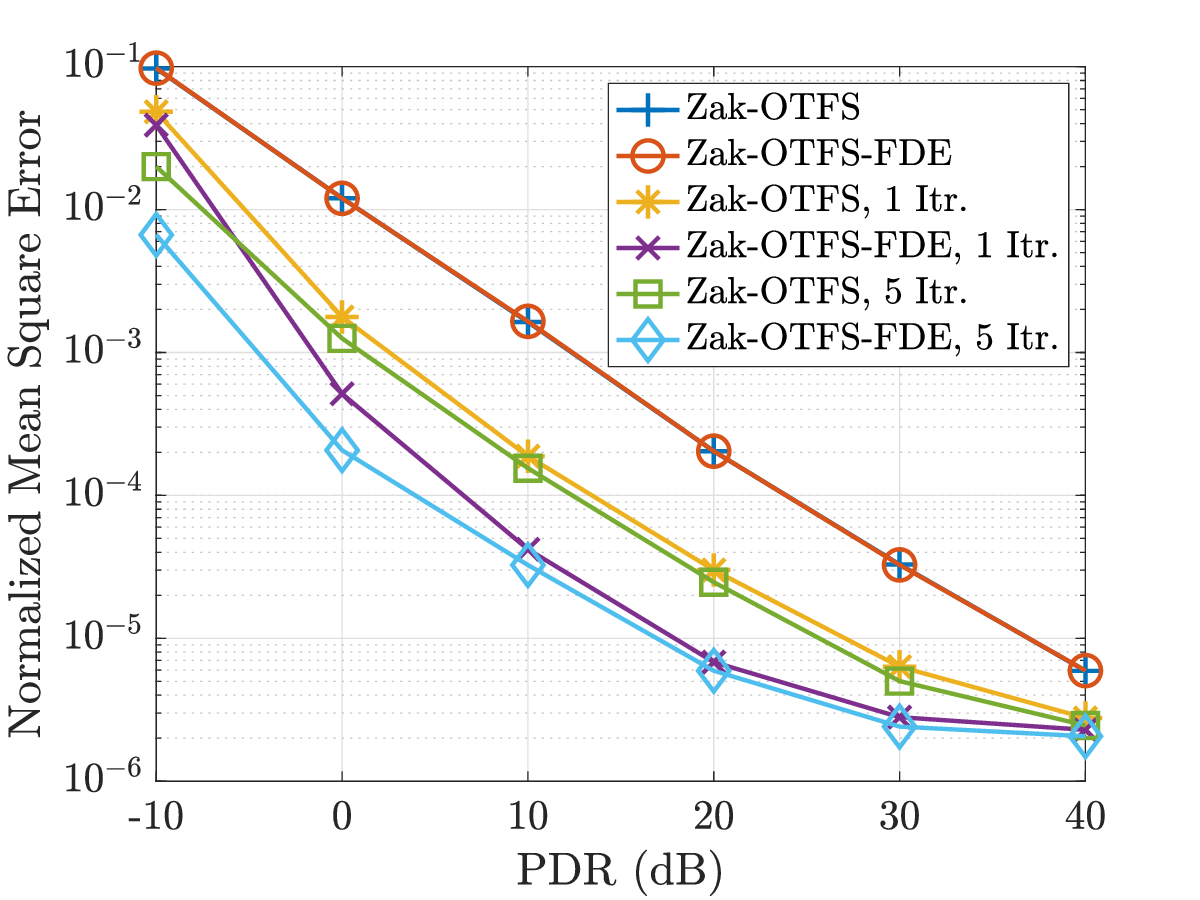}\label{fig:nmse_vs_pdr_rrc}}
    \hfill
    \subfloat[{BER}]{\includegraphics[width=0.45\linewidth]{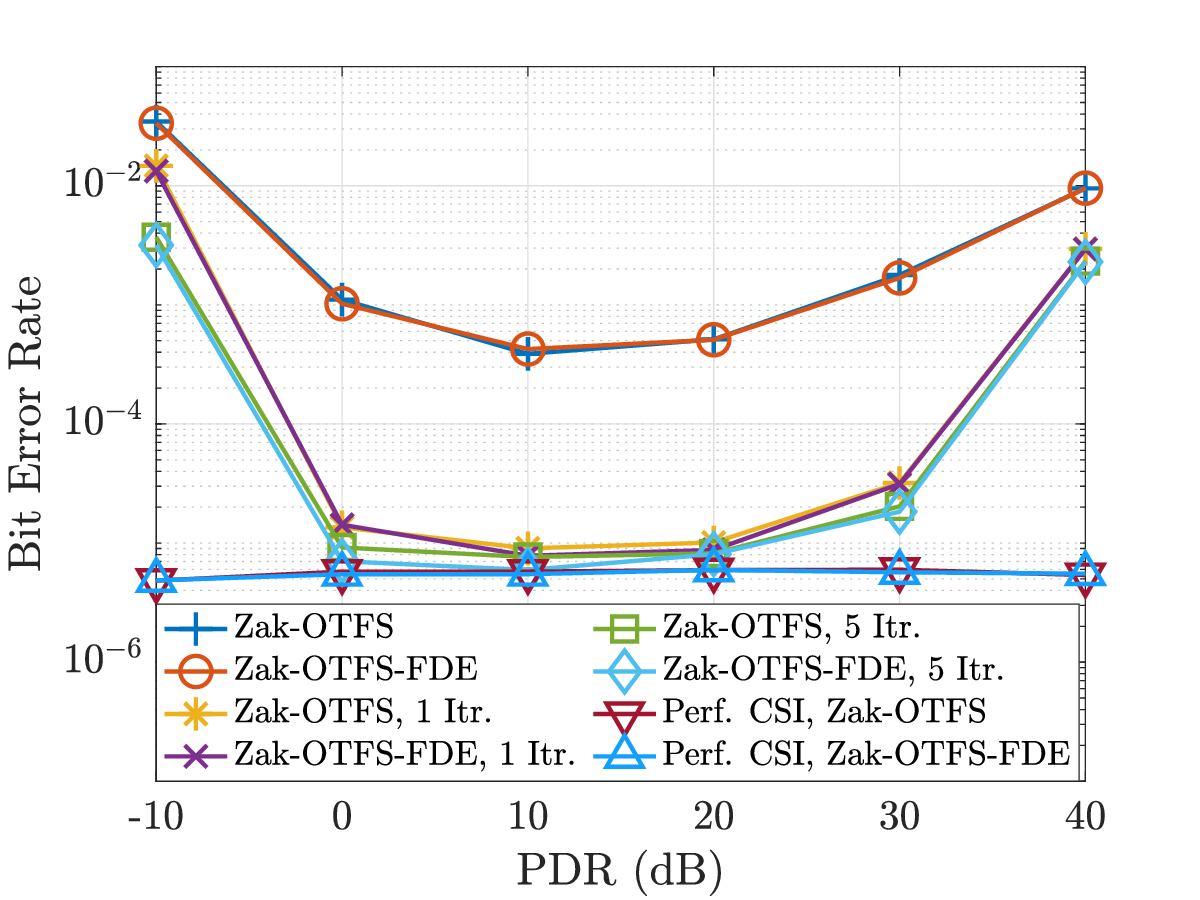}\label{fig:ber_vs_pdr_rrc}}
    \hfill
    \caption{NMSE and BER performance with estimated channel. A Zadoff-Chu sequence with root $101$ in the DD domain \cite{preamblepaper} is used as a spread pilot. FD equalization (FDE) of Zak-OTFS via CGM algorithm with $k=250, \epsilon=10^{-6}, b=\lceil \nu_{\max}T\rceil+1$. Simulation parameters: $M = 31$, $N = 37$, $\nu_{p} = 30$ kHz, data SNR = 20 dB, Veh-A channel with $\nu_{\max} = 815$ Hz, RRC pulse shaping filter with $\beta_\tau = \beta_\nu = 0.6$. NMSE and BER both improve with turbo iterations and achieve close to perfect CSI performance.}
    \label{fig:est_csi_perf_rrc}
\end{figure*}

\begin{figure*}
    \subfloat[{NMSE}]{\includegraphics[width=0.45\linewidth]{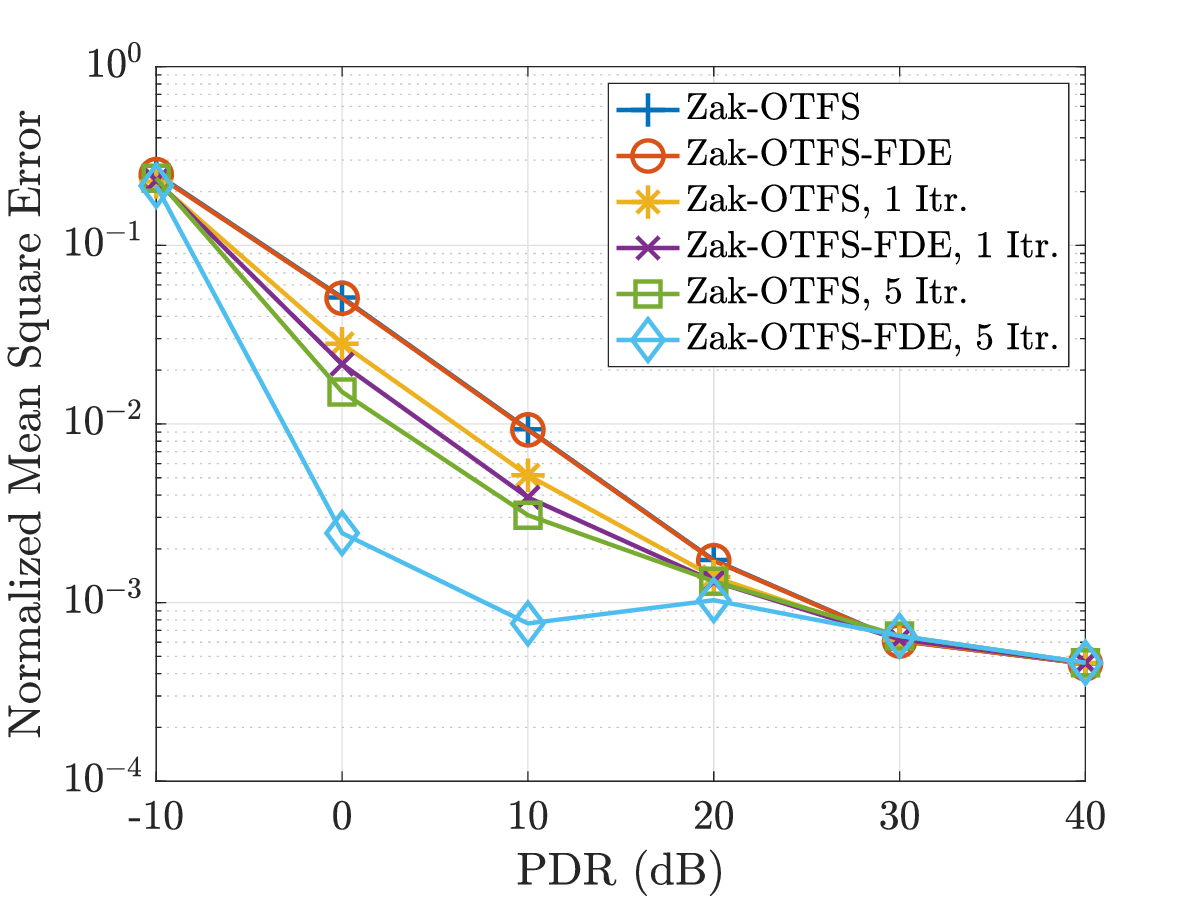}\label{fig:nmse_vs_pdr_gs}}
    \hfill
    \subfloat[{BER}]{\includegraphics[width=0.45\linewidth]{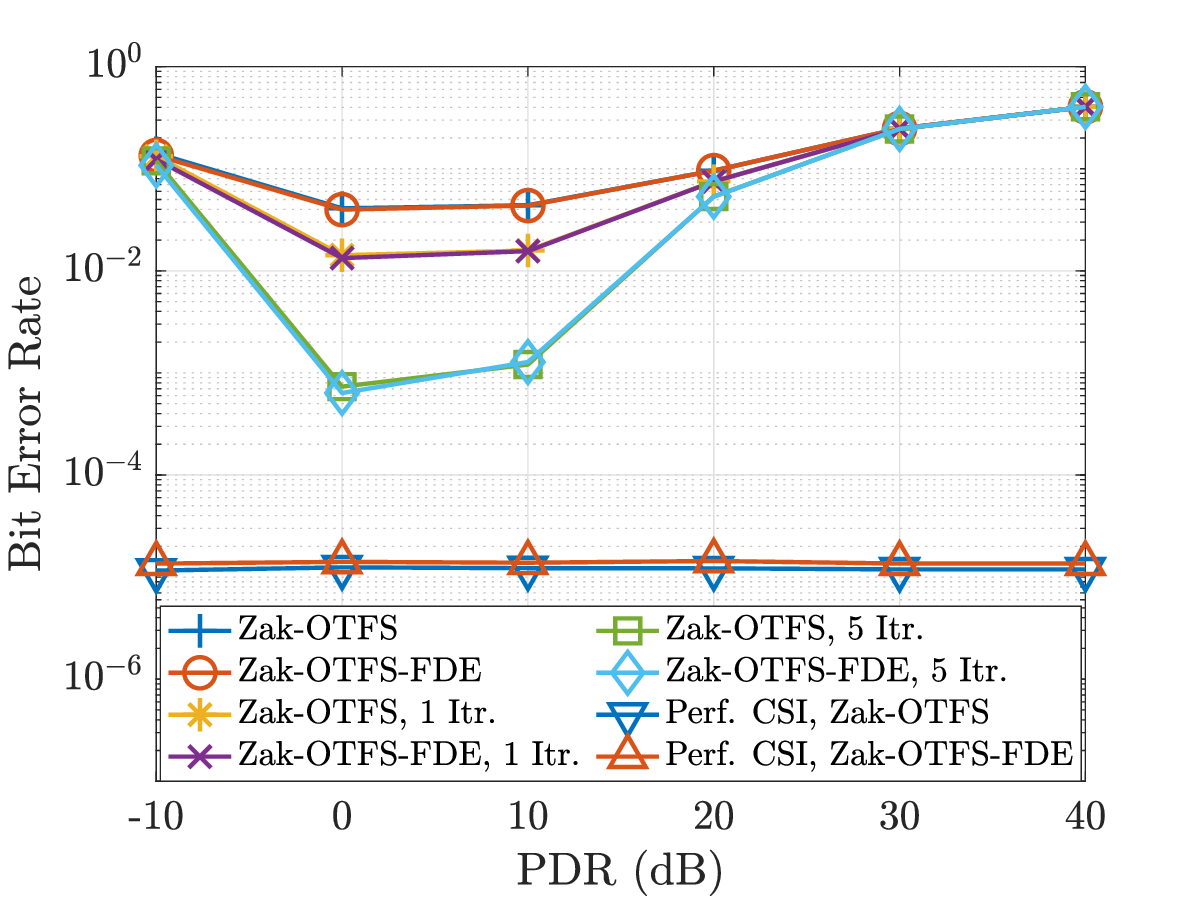}\label{fig:ber_vs_pdr_gs}}
    \hfill
    \caption{NMSE and BER performance with estimated channel. A Zadoff-Chu sequence with root $101$ in the DD domain \cite{preamblepaper} is used as a spread pilot. FD equalization (FDE) of Zak-OTFS via CGM algorithm with $k=250, \epsilon=10^{-6}, b=\lceil 5\nu_{\max}T\rceil$. Simulation parameters: $M = 31$, $N = 37$, $\nu_{p} = 30$ kHz, data SNR = 20 dB, Veh-A channel with $\nu_{\max} = 815$ Hz, Guass-sinc pulse shaping with $\alpha_\tau = \alpha_\nu = 0.044$. NMSE and BER both improve with turbo iterations and achieve close to perfect CSI performance.}
    \label{fig:est_csi_perf_gs}
\end{figure*}

\section{Numerical Results}
\label{sec:results}


\begin{figure}
    \centering
    \includegraphics[width=0.99\linewidth]{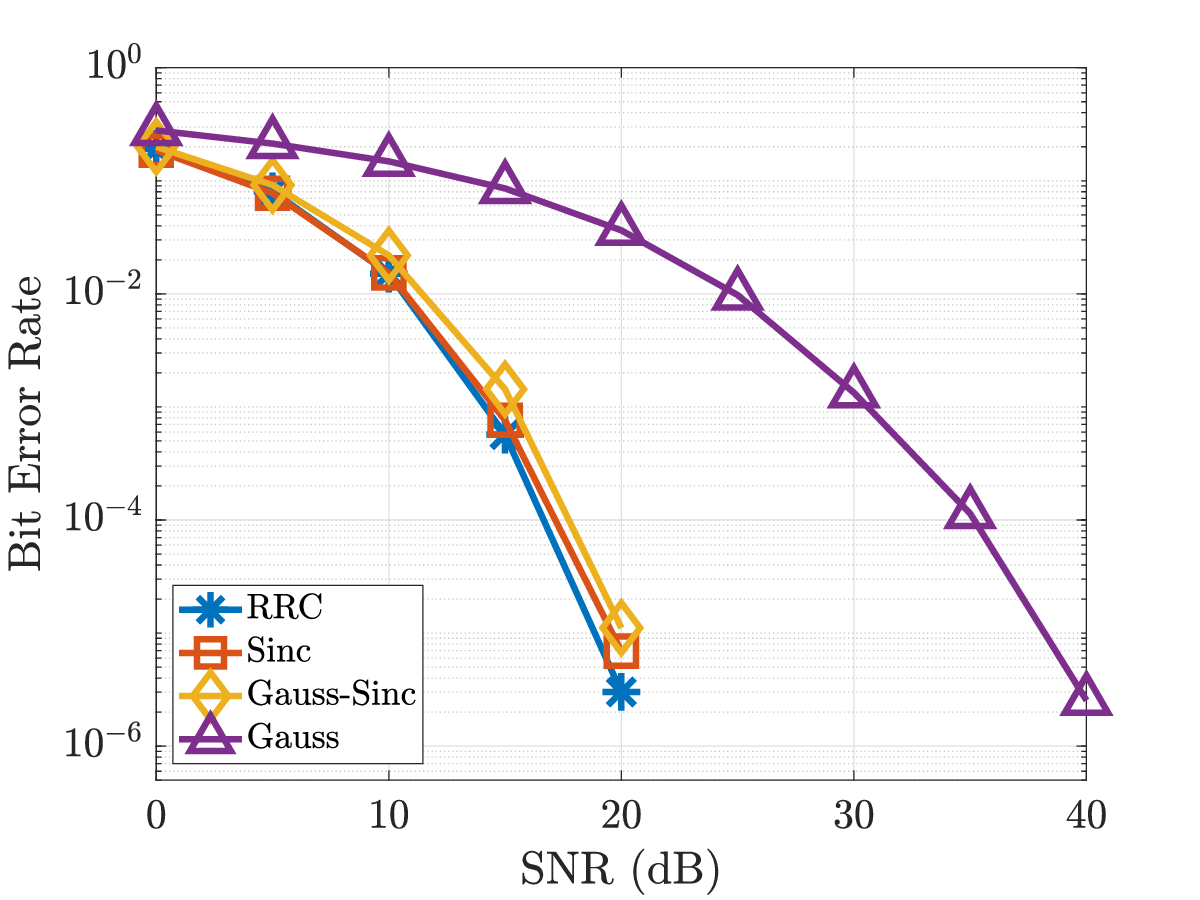}
    \caption{Impact of pulse shaping filter choice on BER of FD equalization (FDE) of Zak-OTFS with perfect channel knowledge. FD equalization of Zak-OTFS via CGM algorithm with $k=250, \epsilon=10^{-6}$. Simulation parameters: $M = 31$, $N = 37$, $\nu_{p} = 30$ kHz, Veh-A channel with $\nu_{\max} = 815$ Hz. For RRC, $b = \lceil \nu_{\max}T\rceil+1$ and $\beta_\tau = \beta_\nu = 0.6$. For sinc, $b = N+1$. For Gauss-sinc, $b = \lceil 5\nu_{\max}T\rceil$ and $\alpha_\tau = \alpha_\nu = 0.044$. For Gauss, $b = \lceil \nu_{\max}T\rceil+1$ and $\alpha_\tau = \alpha_\nu = 1.584$.}
    \label{fig:perf_csi_vs_filter_M31_N37}
\end{figure}

\begin{figure*}
    \subfloat[{NMSE}]{\includegraphics[width=0.45\linewidth]{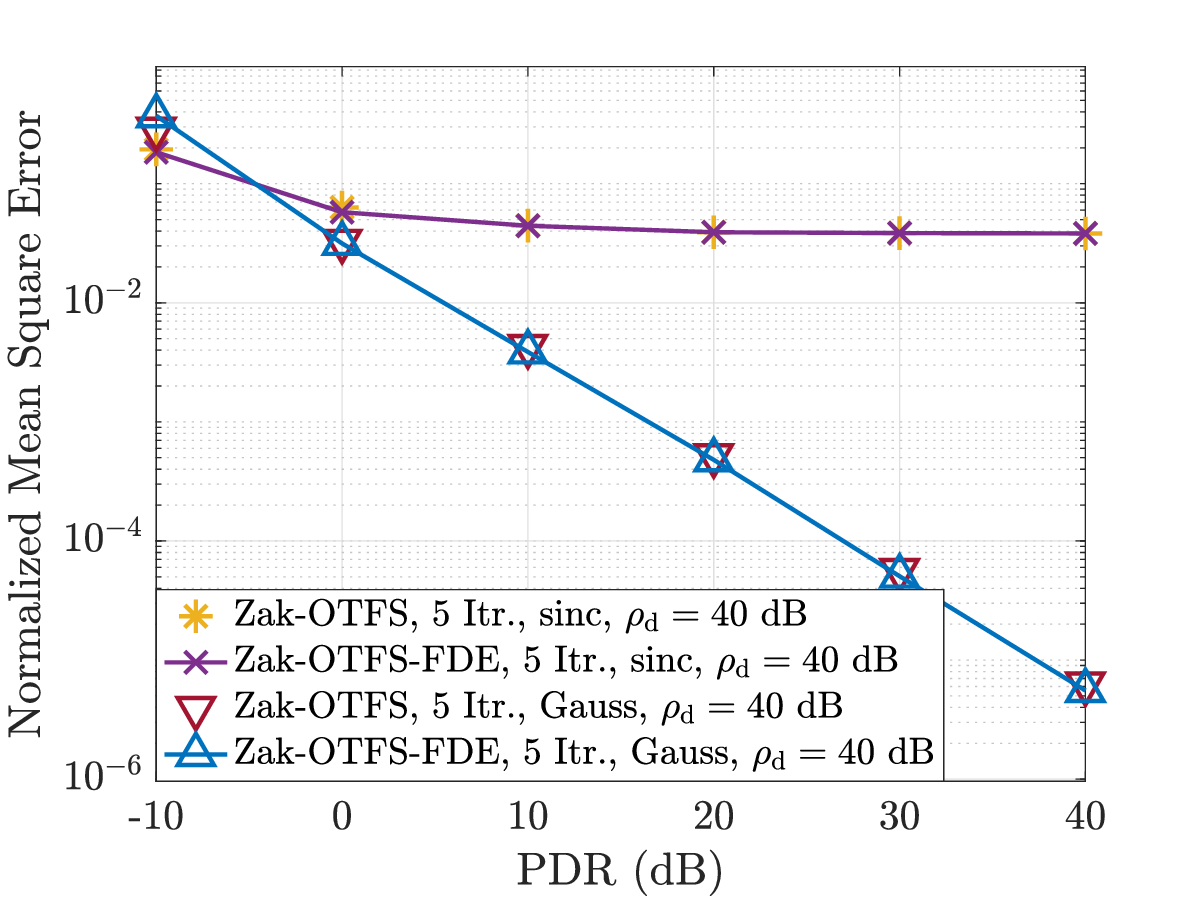}\label{fig:nmse_vs_pdr_guass_and_sinc}}
    \hfill
    \subfloat[{BER}]{\includegraphics[width=0.45\linewidth]{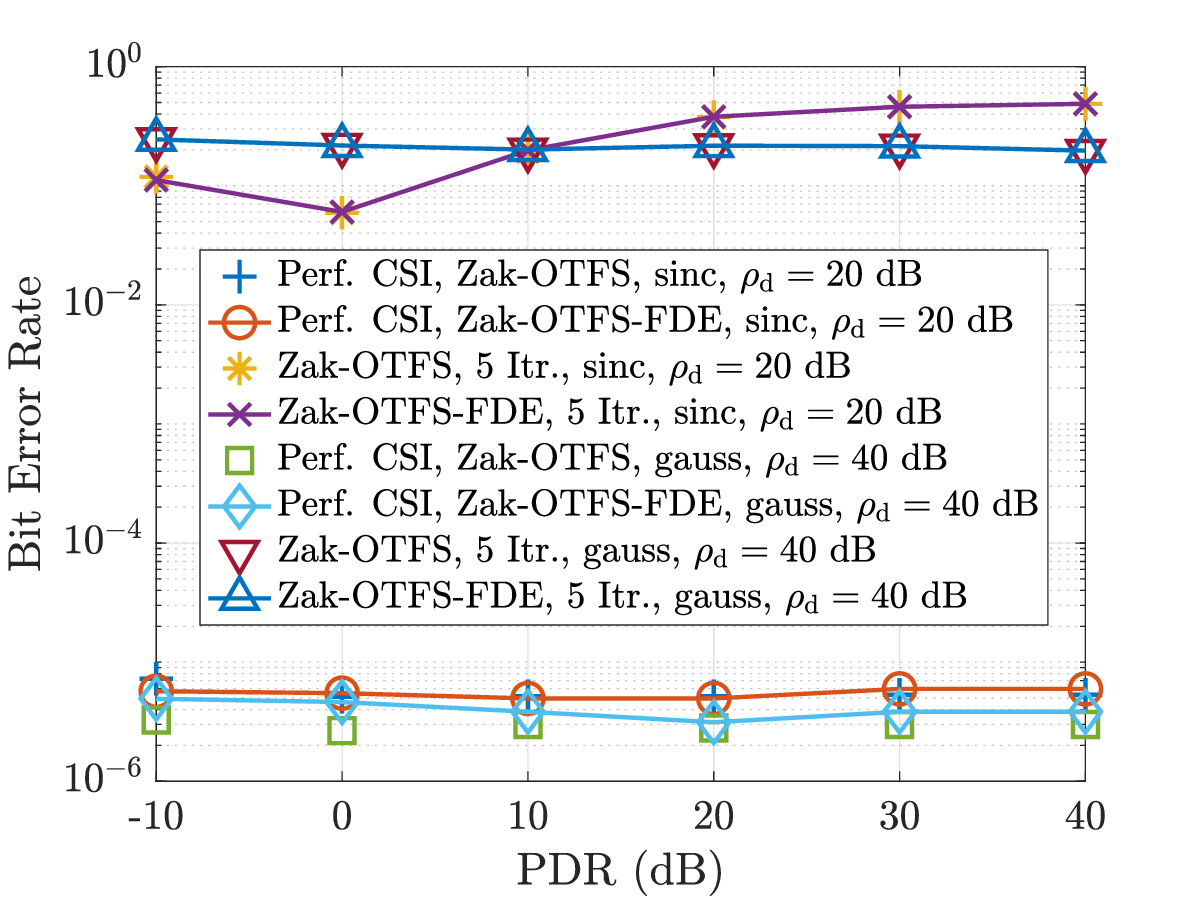}\label{fig:ber_vs_pdr_guass_and_sinc}}
    \hfill
    \caption{NMSE and BER performance with estimated channel. A Zadoff-Chu sequence with root $101$ in the DD domain \cite{preamblepaper} is used as a spread pilot. Sinc pulse shaping and Guass pulse shaping with $a_\tau = a_\nu = 1.584$, $M=31, N=37, \nu_p=30$ kHz, data SNR = 20 dB for sinc and 40 dB for Gauss, $b = N+1$ for sinc and $b = \lceil \nu_{\max}T\rceil+1$ for Gauss. Veh-A channel with $\nu_{\max}=815$ Hz. Sinc has poor NMSE that does not help with the bit-error performance, but Gauss has better NMSE. However, the Gauss pulse is not orthogonal and has poor equalization performance with estimated channel \cite{das2025gaussian}.}
    \label{fig:est_csi_perf_gauss_and_sinc}
\end{figure*}

In this Section, we evaluate the performance of the FD equalization of Zak-OTFS under both perfect channel state information (CSI) and estimated CSI. For all the simulations we consider a Zak-OTFS frame with $M=31, N=37$ and $\nu_p=30$ kHz. We consider a Vehicular-A (Veh-A) channel \cite{veh_a} with power-delay profile shown in Table \ref{tab:veh_a}. Doppler for path $i$ is computed as $\nu_i = \nu_{\max}\cos(\theta_i)$, where $\theta_i$ is uniformly distributed between $-\pi$ and $\pi$ and $\nu_{\max}$ is the maximum Doppler spread\footnote{Note that our channel model considers \textit{fractional} delay and Doppler shifts, which is representative of real propagation environments. The path delays in Table~\ref{tab:veh_a} are non-integer multiples of the delay resolution $\nicefrac{1}{B}$. The Doppler shifts $\nu_i = \nu_{\max}\cos(\theta_i)$ are also non-integer multiples of the Doppler resolution $\nicefrac{1}{T}$ since $\cos(\theta_i)$ is continuous valued.}. We present results for four pulse shapes, sinc~\cite{bitspaper2}, root raised cosine (RRC)~\cite{Aug2024paper, preamblepaper}, Guass~\cite{das2025gaussian}, and Gauss-sinc~\cite{das2025gaussian}. For CP-OFDM, we consider a cyclic prefix of $4$ delay bins. The choice of $b$ depends on the maximum Doppler spread of the channel and the pulse shaping employed. For the simulations presented in the following, the choice of $b$ is specified in the figure captions.

\subsection{Comparison with Existing Approaches}
We compare the perfect CSI performance of Zak-OTFS in DD and FD, I/O relation in~\eqref{eq:mp_ofdm1}, and CP-OFDM in Figs.~\ref{fig:perf_csi_vs_algo_M31_N37_highdop}~and~\ref{fig:perf_csi_vs_algo_M31_N37_lowdop} under high Doppler spread ($\nu_{\max}=815$ Hz) and low Doppler spread ($\nu_{\max} = 81.5$ Hz), respectively. We consider an RRC pulse with parameter $\beta_\tau = \beta_\nu = 0.6$. For the CGM method we consider $k=250, b=\lceil\nu_{\max}T\rceil+1$. For~\eqref{eq:mp_ofdm1} and CP-OFDM we consider two receiver architectures. First a genie receiver which allows joint equalization across carriers (because in practice, it is hard to acquire channel spread occurring due to inter-carrier interference (ICI)) and second a practical receiver, which 
which uses a 1-tap
equalizer (which uses channel estimates along the diagonal and doesn't acquire the ICI). As mentioned in Section~\ref{subsec:min_period_ofdm}, for~\eqref{eq:mp_ofdm1} the subcarrier spacing is $\nicefrac{B}{MN} = 810.81$ Hz, while that for CP-OFDM it is $\nicefrac{B}{M} = 30$ kHz. 

At high Doppler spread (Fig. \ref{fig:perf_csi_vs_algo_M31_N37_highdop}), consequently, the performance of 1-tap equalizer for~\eqref{eq:mp_ofdm1} floors (since Doppler spread is of the order of subcarrier spacing). However, the genie receiver performs much better since ICI is acquired at the receiver. There is no performance difference between the two receivers for CP-OFDM. This is because the subcarrier spacing is large enough to be not be affected by the Doppler spread of 815 Hz. Further, the performance of~\eqref{eq:mp_ofdm1} is better than that of CP-OFDM with genie receiver. This is because CP-OFDM fades more than~\eqref{eq:mp_ofdm1} (see Fig. \ref{fig:column_norm_comp} and the discussion therein). Compared to the OFDM schemes, Zak-OTFS in DD and FD perform much better, thanks to the non-fading aspect for these schemes. The performance of DD and FD equalization is essentially the same, where in DD we used the regular LMMSE incurring complexity $\mathcal{O}(M^3N^3)$ and in FD we used the CGM described in Section~\ref{subsec:low_compl_eq} which has complexity $\mathcal{O}(kbMN)$ (see Section~\ref{subsubsec:complexity}). 

At low Doppler spread (Fig. \ref{fig:perf_csi_vs_algo_M31_N37_lowdop}),~\eqref{eq:mp_ofdm1} and CP-OFDM have similar performance for both receivers. This is expected since the Doppler spread is now $81.5$ Hz and the subcarrier spacing for both CP-OFDM and~\eqref{eq:mp_ofdm1} supports this Doppler spread. As seen before, the performance of Zak-OTFS with LMMSE equalizer in DD and CGM equalizer in FD is similar and better than CP-OFDM and~\eqref{eq:mp_ofdm1}.

\subsection{Performance with Estimated CSI}
Here we present the results of the proposed receiver with estimated channel. For channel estimation, we consider a Zadoff-Chu sequence with root 101 in the DD domain as the spread pilot frame \cite{preamblepaper}. For all the simulations, we fix the data SNR (and data energy) and vary the pilot energy. The ratio of pilot energy to the data energy is 
defined to be the
pilot to data ratio (PDR). To evaluate the channel estimation performance, we compute the normalized mean square error (NMSE) as:
\begin{align}
    \label{eq:NMSE}
    \mathrm{NMSE} = \frac{\Vert \widehat{\mathbf{h}}_{\mathrm{eff}} - \mathbf{h}_{\mathrm{eff}}\Vert^2_2}{\Vert\mathbf{h}_{\mathrm{eff}}\Vert_2^2},
\end{align}
where $\mathbf{h}_{\mathrm{eff}}$ is the effective channel matrix (see \eqref{eq:sys_model1}) and $\widehat{\mathbf{h}}_{\mathrm{eff}}$ is the estimated effective channel matrix using the spread pilot.

In Figs.~\ref{fig:est_csi_perf_rrc}~and~\ref{fig:est_csi_perf_gs} we
compare the NMSE and BER performance of Zak-OTFS in DD and FD with estimated channel 
for RRC and Gauss-sinc pulse shapes, respectively. For Gauss-sinc pulse shape we fix the parameters to be $\alpha_\tau=\alpha_\nu=0.044$~\cite{das2025gaussian} and the corresponding CGM parameters to be $b=\lceil5\nu_{\max}T\rceil$. The NMSE and BER performance with turbo iterations (described in Section~\ref{subsec:turbo_iter}) is also added. 

For RRC pulse shaping in Fig.~\ref{fig:est_csi_perf_rrc}\subref{fig:nmse_vs_pdr_rrc}, the NMSE performance without turbo iterations decreases linearly with PDR for FD and DD equalization of Zak-OTFS. This is expected since increasing PDR corresponds to increasing the pilot energy. With 1 turbo iteration, the NMSE performance is seen to improve with the improvement being more significant with 5 turbo iterations. The performance of both FD and DD equalization of Zak-OTFS improves similarly with turbo iterations.
BER performance for the RRC pulse is plotted in Fig.~\ref{fig:est_csi_perf_rrc}\subref{fig:ber_vs_pdr_rrc}. Performance assuming perfect CSI is also added for comparison. As observed before in Fig. \ref{fig:perf_csi_vs_algo_M31_N37_highdop}, the perfect CSI performance for both FD and DD equalizations match and is flat across PDR, since perfect CSI allows perfect pilot cancellation (see \eqref{eq:ch_est_4}). Without turbo iterations, the BER performance is first seen to decrease with PDR and then increase. This can be explained as follows. At very low PDR, the channel estimate has high NMSE and therefore leads to poor equalization and detection performance. As the PDR increases, the accuracy of the channel estimate improves, which results in better BER performance. However, as the PDR is further increased, although the NMSE improves, the energy of the residual pilot left behind after the pilot cancellation step \eqref{eq:ch_est_4} increases and this causes increased interference during data detection. Further since at each step the pilot cancellation is not exact, the BER performance is far from perfect CSI performance. However, with turbo iterations, due to repeated pilot and data cancellation steps, the performance significantly improves and achieves the perfect CSI performance for mid PDR values. Notice that with and without turbo, the performance of both FD and DD equalization of Zak-OTFS is similar. However, the NMSE in FD is slightly better than that in DD because of decreased interference between information symbols and pilot symbols (recall that there are only $MN-2b$ information symbols in FD compared to $MN$ information symbols in DD). NMSE and BER results plotted in Figs.~\ref{fig:est_csi_perf_gs}\subref{fig:nmse_vs_pdr_gs} and~\ref{fig:est_csi_perf_gs}\subref{fig:ber_vs_pdr_gs}, respectively, for Gauss-sinc pulse shaping also show a similar trend as in the case of RRC pulse shaping. 

\subsection{Impact of Pulse Shaping Filter}
Here we present the simulation results comparing different pulse shapes with perfect and estimated CSI. Fig.~\ref{fig:perf_csi_vs_filter_M31_N37} shows the performance of Zak-OTFS in FD using RRC, sinc, Gauss-sinc, and Gauss pulse shapes with perfect CSI. The CGM method in Algorithm~\ref{alg:conj_grad} is used for equalization. The performance of sinc, RRC, and Gauss-sinc are observed to be similar. Gauss pulse shaping has degraded performance owing to its non-orthogonality \cite{das2025gaussian}.

Fig.~\ref{fig:est_csi_perf_gauss_and_sinc} shows the estimated CSI performance of both Zak-OTFS in FD and DD for sinc and and Gauss pulse shapes. The NMSE performance of sinc in Fig.~\ref{fig:est_csi_perf_gauss_and_sinc}\subref{fig:nmse_vs_pdr_guass_and_sinc} is seen to floor. This is because, the sinc pulse decays slowly and there is significant energy outside the fundamental rectangle that is not estimated \cite{das2025gaussian}. On the other hand, the Gauss pulse has excellent NMSE performance since it is localized within the fundamental and the NMSE decreases linearly with PDR. The BER performance of Zak-OTFS in FD and DD for sinc and Gauss pulse shapes is plotted in Fig.~\ref{fig:est_csi_perf_gauss_and_sinc}\subref{fig:ber_vs_pdr_guass_and_sinc}. The perfect CSI performance is also plotted for comparison. The estimated CSI performance of sinc for both FD and DD equalization is far from perfect CSI performance. This is due to the poor NMSE achieved by the sinc pulse shape. Turbo iterations also do not help since, even with turbo, we estimate the channel only within the fundamental rectangle. Gauss pulse shape also has poor estimated CSI performance. This is attributed to non-orthogonality of the Gauss pulse shape that leads to poor equalization~\cite{das2025gaussian}. Again, for all these cases, the FD and DD equalization performances match, showing its equivalence under perfect CSI, estimated CSI, and different pulse shapes.

\subsection{Comparison with AFDM~\cite{afdm_1, afdm_2}}
\label{subsec:afdm_vs_zak}

\begin{figure*}
    \subfloat[{$MNc_1 = 91.2 \not\in \mathbb{Z}$}]{\includegraphics[width=0.45\linewidth]{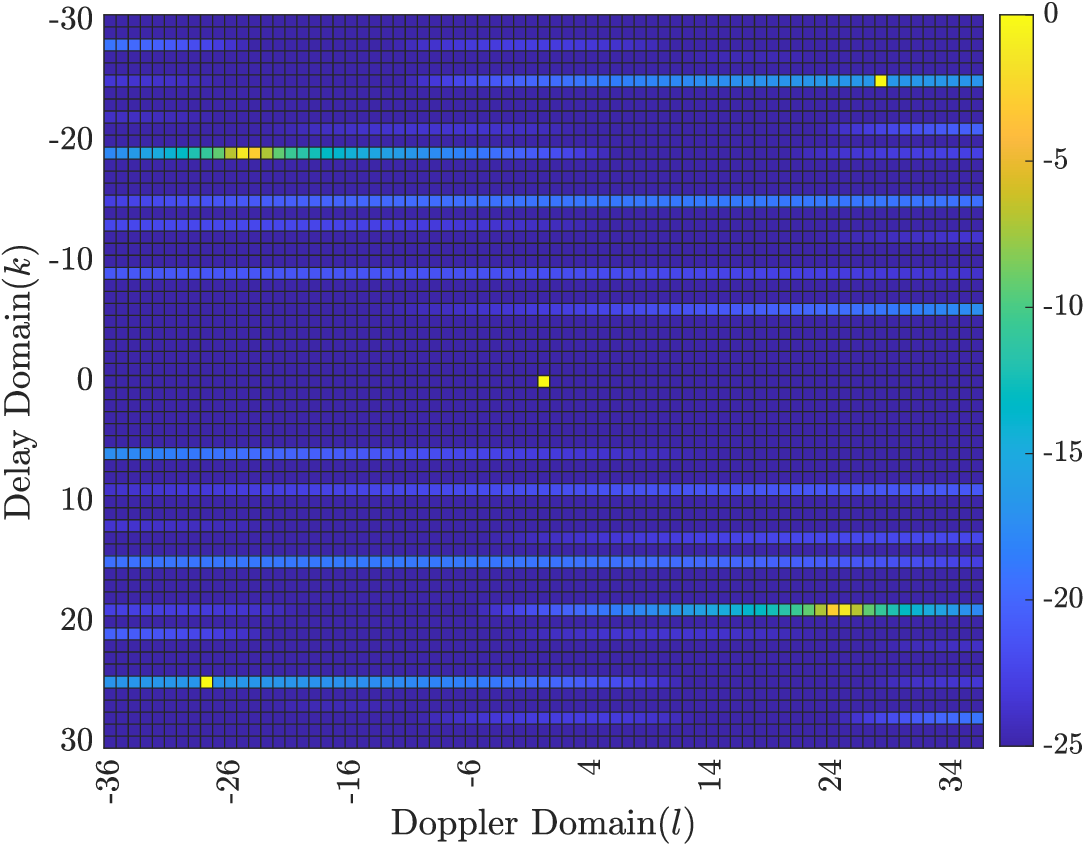}\label{fig:afdm_sa_91p2}}
    \hfill
    \subfloat[{$MNc_1 = 91 \in \mathbb{Z}$}]{\includegraphics[width=0.45\linewidth]{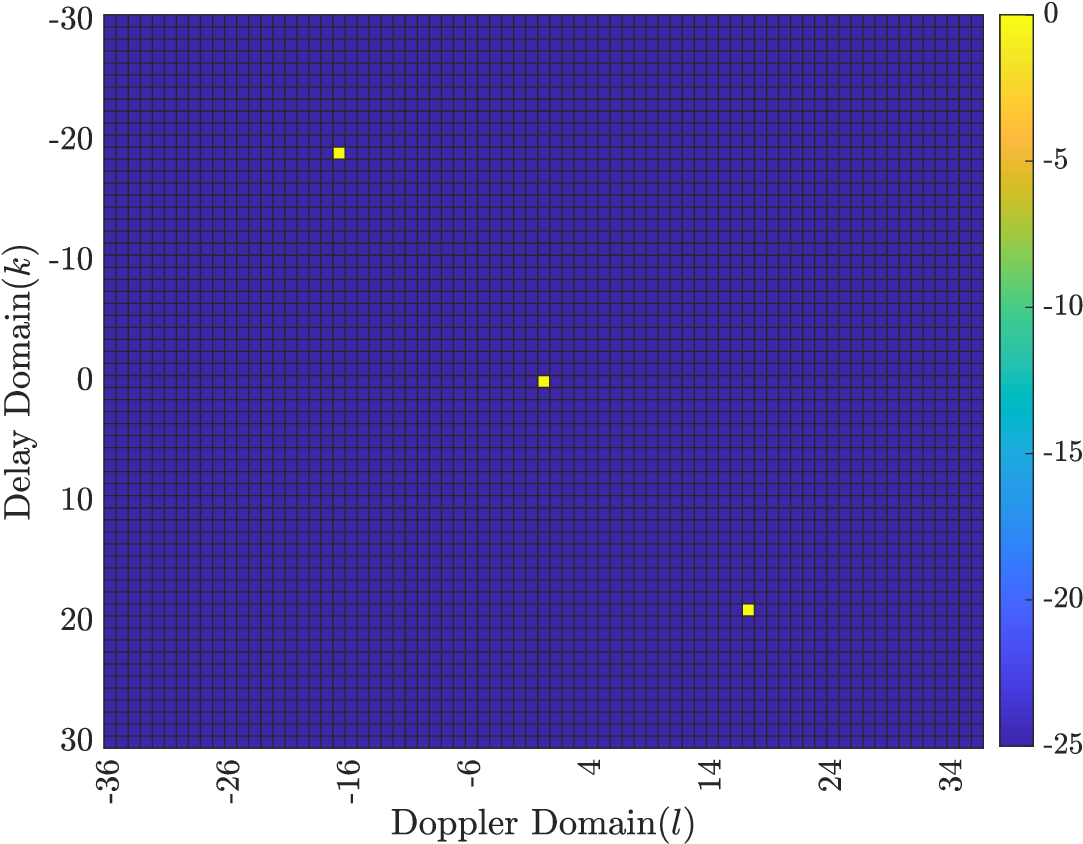}\label{fig:afdm_sa_91}}
    \hfill
    \caption{Heatmap of self-ambiguity of basis function of AFDM in dB scale. $M=31, N=37, m=0$.}
    \label{fig:sa_comp_otfs_afdm}
\end{figure*}

The basis used to modulate information symbols in the transfer domain in AFDM is~\cite{afdm_1}:
\begin{align}
    \label{eq:afdm1}
    \boldsymbol{\phi}_m[n] = \frac{1}{\sqrt{MN}}e^{j2\pi\big(c_1n^2+c_2m^2+\frac{mn}{MN}\big)},
\end{align}
where $m, n = 0, 1, \cdots, MN-1, c_1, c_2 \in \mathbb{R}$. $n$ is the time index and $m$ is the transfer domain index. Substituting in~\eqref{eq:sys_model5}, the received symbols in the transfer domain is:
\begin{align}
    \label{eq:afdm2}
    \breve{\mathbf{r}}[f] &= \frac{1}{MN}\sum_{i = 0}^{MN-1} \breve{\mathbf{s}}[i] \bigg(\sum_{\bar{k}=0}^{MN-1} \sum_{\bar{l}=0}^{MN-1} e^{-\frac{j2\pi}{MN}\bar{l}\bar{k}} \mathbf{h}[\bar{k},\bar{l}] \times \nonumber \\ &~~~\sum_{n=0}^{MN-1} e^{c_2(f^2-i^2)+n\big(\frac{i-f}{MN}-2c_1\bar{k}\big)-\frac{i\bar{k}}{MN}+c_1\bar{k}^2}\bigg) \nonumber \\
    &= \sum_{i=0}^{MN-1}\breve{\mathbf{s}}[i]\breve{\mathbf{H}}[f, i],
\end{align}
and the system model is:
\begin{align}
    \label{eq:afdm3}
    \breve{\mathbf{r}} = \breve{\mathbf{H}}\breve{\mathbf{s}} + \breve{\mathbf{n}},
\end{align}
where $\breve{\mathbf{s}}, \breve{\mathbf{r}}, \breve{\mathbf{n}} \in \mathbb{C}^{MN\times 1}$ denote the vector of transmitted symbols, received symbols and noise samples, respectively, all in the transfer domain. Since $\mathbf{h}[\bar{k},\bar{l}]$ encompasses pulse shaping and fractional DD values, the system model in \eqref{eq:afdm3} also holds for various pulse shapes and fractional DD values. 

\subsubsection{Choices of $c_1$ and $c_2$}
There are two ways of choosing $c_1$ and $c_2$: $(i)$ $c_1MN,~c_2MN \in \mathbb{Z}$, or $(ii)$ $c_1MN,~c_2MN \not\in \mathbb{Z}$.\\
\noindent\textit{Case-I:} $c_1MN,~c_2MN \in \mathbb{Z}$. The basis can be represented as:
\begin{align}
    \label{eq:afdm4}
    \boldsymbol{\phi}'_m[n] = \frac{1}{\sqrt{MN}}e^{j2\pi\big(\frac{\alpha}{MN}n^2+\frac{\beta}{MN}m^2+\frac{mn}{MN}\big)},
\end{align}
for some $\alpha, \beta \in \mathbb{Z}$.
\begin{lemma}
    \label{lemma:afdm_cazac}
    $\boldsymbol{\phi}'_m[n]$ is a constant amplitude zero-autocorrelation (CAZAC) sequence.
\end{lemma}
\begin{IEEEproof}
    Clearly $\boldsymbol{\phi}'_m[n]$ is constant amplitude sequence. Consider the inner product:
    \begin{align}
        \label{eq:afdm5}
        \langle\boldsymbol{\phi}'_m[n], \boldsymbol{\phi}'^*_m[n-k]\rangle &= \frac{1}{MN}\sum_{n=0}^{MN-1}e^{j2\pi\big(\frac{\alpha}{MN}n^2+\frac{\beta}{MN}m^2+\frac{mn}{MN}\big)}\times\nonumber\\
        &\hspace{5mm}e^{-j2\pi\big(\frac{\alpha}{MN}(n-k)^2+\frac{\beta}{MN}m^2+\frac{m(n-k)}{MN}\big)} \nonumber \\
        &= \frac{1}{MN}\sum_{n=0}^{MN-1}e^{-j2\pi\big(\frac{\alpha}{MN}(k^2-2nk)-\frac{km}{MN}\big)} \nonumber \\
        &\overset{(a)}{=} \begin{cases}
            1 \quad \text{if } \ 2k \equiv 0 \bmod{MN} \\
            0 \ \ \quad \text{otherwise}
        \end{cases},
    \end{align}
    where step $(a)$ follows from Identity~\ref{idty:sumrootsofunity}.
\end{IEEEproof}
Conventionally, Zak-OTFS is realized through the pulsone basis. Zak-OTFS can also be realized through CAZAC basis~\cite{mehrotra2025Zak} and both these realizations are unitarily equivalent. For the values of $c_1$ and $c_2$ in Case-I, AFDM bases are CAZAC sequences. Hence, for this choice of $c_1$ and $c_2$, Zak-OTFS and AFDM are unitarily equivalent. Note that, prior papers on AFDM~\cite{afdm_1, afdm_2} choose $c_1MN$ to be integer-valued.\\
\noindent\textit{Case-II:} $c_1MN,~c_2MN \not\in \mathbb{Z}$. In this cases the basis is no longer a CAZAC sequence, since the exponent cannot be written as roots of unity. For the values of $c_1$ and $c_2$ in Case-II, AFDM and Zak-OTFS are not unitarily equivalent.

\begin{figure}
    \centering
    \includegraphics[width=0.99\linewidth]{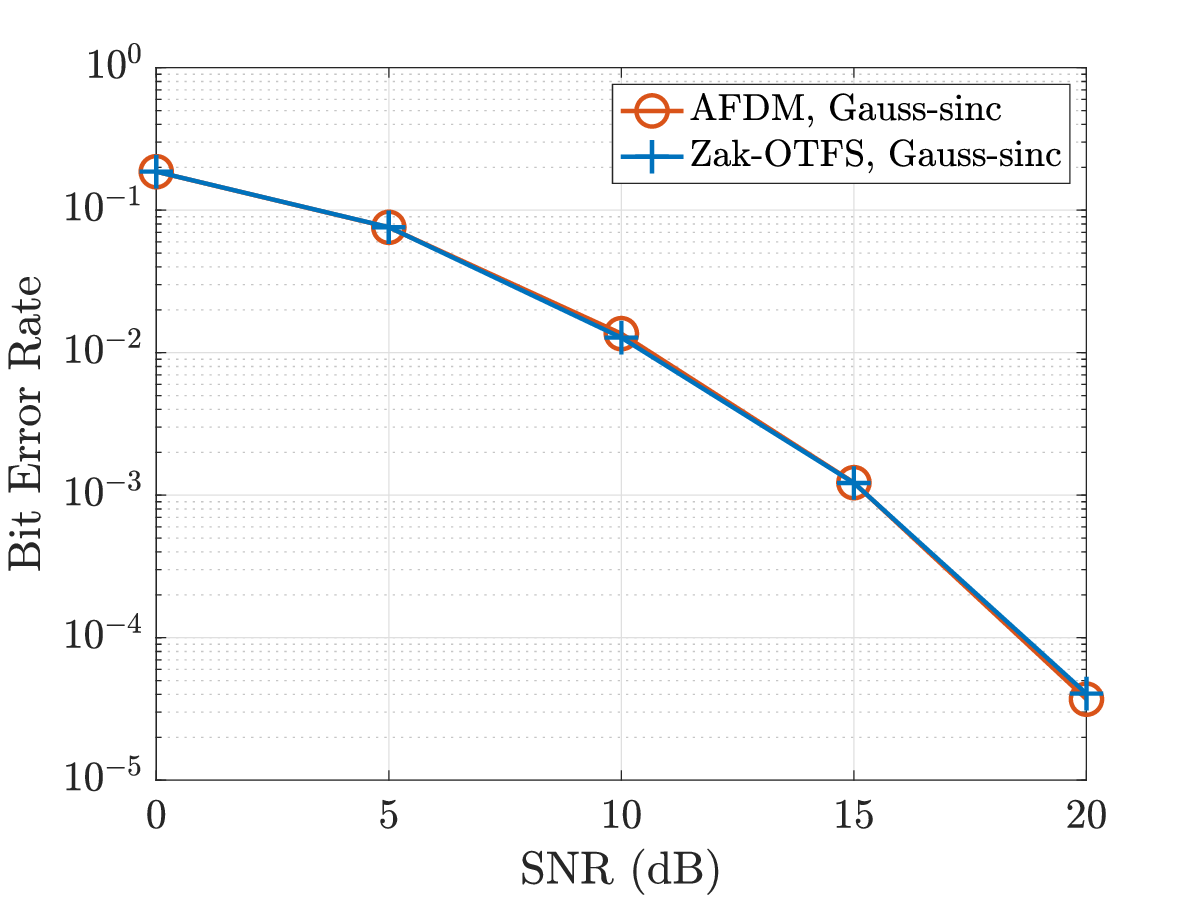}
    \caption{BER performance comparing the performance of Zak-OTFS and AFDM with $c_1MN, c_2MN \in \mathbb{Z}$ for Gauss-sinc pulse shaping with $\alpha_{\tau} = \alpha_{\nu} = 1.584, M = 11, N = 13, \nu_p = 30$ kHz. Veh-A channel with $\nu_{\max}=815$ Hz. There is essentially no difference between AFDM and Zak-OTFS.}
    \label{fig:ber_comp_otfs_afdm}
\end{figure}

Fig.~\ref{fig:sa_comp_otfs_afdm}\subref{fig:afdm_sa_91p2} shows the heatmap of the absolute value of the self-ambiguity function when $c_1$ is chosen to be a value in Case-II, i.e., $c_1MN = 91.2 \not\in\mathbb{Z}$ and $m=0$. Fig.~\ref{fig:sa_comp_otfs_afdm}\subref{fig:afdm_sa_91} shows the heatmap when $c_1$ is chosen to be a value in Case-I, i.e., $c_1MN = 91 \in\mathbb{Z}$ and $m=0$. For estimating the channel in the DD domain it is desirable to have extremely localized self-ambiguity function and this is possible only when $c_1$ is chosen to be a value in Case-I, i.e., when AFDM and Zak-OTFS are unitarily equivalent. AFDM can be regarded as a general case of Zak-OTFS, however, the generalization beyond Zak-OTFS makes channel estimation more challenging. Fig.~\ref{fig:ber_comp_otfs_afdm} compares the BER performance of AFDM and Zak-OTFS under perfect channel knowledge when $c_1$ is chosen to be a value in Case-I. The BER curves overlap corroborating the above results.

\section{Conclusion}
\label{sec:conclusion}
In this paper, we proposed a low-complexity equalization scheme for Zak-OTFS in the frequency domain. We derived the frequency-domain system model and showed that it is unitarily equivalent to the conventional delay-Doppler system model. The frequency-domain Zak-OTFS channel matrix was shown to have a modulo banded structure. We reduced the matrix to a banded structure and leveraged this structure in the CGM algorithm to achieve low-complexity equalization. We showed that the proposed scheme achieves complexity $\mathcal{O}(bkMN)$ that is linear in $M$ and $N$ while that for conventional Zak-OTFS schemes is $\mathcal{O}(M^3N^3)$. Through numerical simulations we verified that equalization of Zak-OTFS in FD has similar performance as DD equalization under both perfect and estimated CSI across pulse shaping filters.

\bibliographystyle{IEEEtran}
\bibliography{references}

\end{document}